%% file: bare_jrnl_new_sample4.tex
\documentclass[journal]{IEEEtran}
\usepackage[T1]{fontenc}
\usepackage{amsmath,amsfonts}
\usepackage{algorithmic}
\usepackage{algorithm}
\usepackage{array}
\usepackage[caption=false,font=normalsize,labelfont=sf,textfont=sf]{subfig}
\usepackage{textcomp}
\usepackage{stfloats}
\usepackage{url}
\usepackage{verbatim}
\usepackage{graphicx}
\usepackage{cite}
\usepackage{xcolor}

\usepackage{tikz}
\usepackage{forest}
\usetikzlibrary{arrows.meta,positioning}
\definecolor{darkblue}{RGB}{56,84,146}
\usepackage{booktabs}

\begin{document}

\title{LLM-Assisted Software Migration: from Module-Level Transformation to Repository-Level Integration}

\author{\author{Yanan Zhang\textsuperscript{1},
Shengqi Liu\textsuperscript{1},
Tao Zhang\textsuperscript{1},
Yidi Liu\textsuperscript{1},
Alan Tang\textsuperscript{1},
Hailong Chen\textsuperscript{2},
and Yantao Jia\textsuperscript{1,*}%
\thanks{\textsuperscript{1}Huawei Technologies Co., Ltd.}%
\thanks{\textsuperscript{2}University of Groningen.}%
\thanks{\textsuperscript{*}Corresponding author.}%
}}

\maketitle

\begin{abstract}
\input{sections/00-abstract}
\end{abstract}


\input{sections/introduction}
\input{sections/background}
\input{sections/methodology}
\input{sections/se_perspective}
\input{sections/role_perspective}
\input{sections/agent_perspective}
\input{sections/enabling-tech}
\input{sections/evaluation}
\input{sections/challenges}

\input{sections/conclusion}

\bibliographystyle{IEEEtran}
\bibliography{references/reference}
\end{document}

%% file: sections/00-abstract.tex
Software migration adapts an existing implementation to a new language, version, API, framework, or platform while preserving its behavior. Large language models (LLMs) succeed on individual functions and modules, but this success does not readily carry over to entire repositories, particularly deployable applications. This survey reviews LLM-assisted migration from module-level transformation to repository-level integration, culminating in the delivery of complete applications. We analyze 96 sources---including 46 direct migration papers---across four perspectives: (1) supported software engineering activities, (2) model roles as editors or orchestrators, (3) orchestration workflows (agentless, single-agent, or multi-agent), and (4) enabling techniques including knowledge engineering, project context, environment feedback, and model adaptation. We then compare evaluation practices and benchmarks. Although 30 of the 46 studies target repositories, most acceptance evidence relies merely on build success, supplied tests, or static and model-based proxies. Moreover, compilation, visual similarity, test outcomes, and runtime behavior capture fundamentally distinct properties and cannot be conflated into a single success rate. We also identify critical gaps in requirements engineering support and in controlled comparisons of agent architectures. Finally, we discuss six open problems: application integration, platform differences, non-functional properties, generalization, attribution of gains, as well as cost and maintenance.

%% file: sections/introduction.tex
\section{Introduction}
\label{sec:introduction}

Software migration adapts an existing implementation to a new target, such as another programming language, a newer language or library version, a different framework, or another platform, while preserving the behavior. 
It is a routine part of software evolution. 
Migration tasks include translating legacy COBOL systems into C for maintainability~\cite{DBLP:journals/corr/abs-2607-04092}, rewriting C code in Rust to obtain memory safety~\cite{DBLP:conf/iceccs/CaiLHYWLWYJ25}, upgrading Java~8 projects to later long-term-support releases~\cite{DBLP:conf/kdd/LiuLZCLNOSHTD26}, replacing deprecated APIs after dependency updates~\cite{DBLP:journals/tse/MahmudDCYNY26, DBLP:journals/ese/ReyesMBNBM26}, and porting mobile applications from Android to HarmonyOS or iOS~\cite{DBLP:journals/corr/abs-2608-09273,DBLP:journals/corr/abs-2507-16037}. 
Despite its necessity, such work is exceedingly costly: large migrations are labor-intensive, offer limited professional incentives for developers, and can span years to complete~\cite{DBLP:conf/sigsoft/ZiftciNSKCK25}.

At first glance, software migration resembles natural-language translation: both map content from a source to a target representation while preserving the underlying semantics. However, this resemblance diverges in three critical aspects. First, the criterion for correctness is far more stringent. While a translated sentence remains functional if its meaning is approximately correct, migrated code is subject to the rigors of a compiler and execution environment. A single misplaced token can break the entire build; thus, behavioral preservation must be verified by test suites or execution oracles rather than mere manual inspection. 
Second, the relevant context for software migration extends far beyond the immediate unit of translation. The correct mapping of a function often depends on types, call sites, and dependencies residing in disparate files, as well as specific build configurations. Consequently, a function translated in isolation may appear syntactically plausible yet fail to integrate with the broader system. 
Third, the target of migration is a holistic ecosystem rather than just a language. It encompasses specific APIs, versioning constraints, and platform contracts (e.g., UI lifecycles and permission models) for which source constructs may lack direct counterparts. Literal translation often yields non-idiomatic results: for instance, rule-based transpilers like C2Rust aim to preserve semantics but emit pervasively unsafe Rust code~\cite{DBLP:journals/pacmpl/EmreSDH21}. 
These fundamental differences explain the limitations of traditional migration tools. Although rule-based transpilers are precise, they must be painstakingly engineered for each specific source--target pair~\cite{DBLP:conf/iclr/LeeMKC0CWR24}, and their output often requires substantial manual refactoring to meet maintainability standards.

Large language models (LLMs) reshape this trade-off. Trained on large corpora of code and documentation, they encode extensive knowledge of diverse languages, frameworks, and platform APIs, extending their utility across a broad spectrum of scenarios. LLMs have already demonstrated success in specialized migration tasks~\cite{zheng2026portingdeclarativeuiharmonyos, DBLP:conf/internetware/GongWCHW25}
with encouraging industrial outcomes. For instance, in a year-long program at Google, an LLM generated 74.45\% of the submitted migration changes~\cite{DBLP:conf/sigsoft/ZiftciNSKCK25}. However, this efficacy diminishes as the migration scope expands beyond isolated components. The best configuration reported on RepoTransBench succeeds on only 32.8\% of repository-level tasks~\cite{DBLP:journals/tse/WangWWGCGLMMZZ26}, and practitioners find LLMs far more acceptable for single-file migrations than for entire projects (62.3\% versus 4.9\%)~\cite{chand_et_al:LIPIcs.ESEM.2026.85}. 
In essence, success at the level of individual functions and modules does not scale to the holistic migration of complete applications, where inter-module dependencies and platform contracts become paramount.

This gap arises not only from the scale of the codebase but fundamentally from the criteria for correctness: while an isolated function can be verified against discrete input--output pairs, a migrated application must build cleanly, execute reliably, and preserve runtime behavior on the target platform. 
This survey focuses on the transition from module-level migration (transforming individual functions, files, API usages, or UI components) to repository-level integration, which delivers a working codebase and, ultimately, a deployable application. 
Research has progressively evolved along this path---from module-level transformation~\cite{DBLP:journals/tse/MahmudDCYNY26, DBLP:conf/internetware/GongWCHW25, zheng2026portingdeclarativeuiharmonyos} to repository-level integration~\cite{DBLP:journals/pacmse/IbrahimzadaKPAPSJ25, DBLP:journals/corr/abs-2604-07341, DBLP:conf/kdd/LiuLZCLNOSHTD26} and fully operational applications~\cite{DBLP:journals/corr/abs-2605-06754, DBLP:journals/corr/abs-2608-09273}. 
Each successive tier introduces complex constraints, such as dependency ordering, build configurations, and platform services, that lower levels fail to verify. 
Mobile ecosystems make these constraints acutely visible. For instance, Android-to-iOS migration often breaks across method, file, and package granularities even with retrieved project context~\cite{DBLP:journals/corr/abs-2507-16037}, while Android-to-OpenHarmony test migration is constrained primarily by platform-specific semantics~\cite{DBLP:journals/corr/abs-2607-11245}. 
We therefore use mobile app migration as a running example, while surveying migration across diverse programming languages, versions, APIs, frameworks, and platforms.

Despite the practical significance of this repository-level transition, existing surveys examine only adjacent topics. 
Specifically, a systematic review of neural code translation~\cite{DBLP:journals/corr/abs-2505-07425} compares translation techniques, code granularities, and evaluation metrics across file- and project-level tasks; however, it does not organize the literature around the end-to-end deliverables required for a complete migration. 
Broader surveys on LLMs for software engineering~\cite{DBLP:journals/tosem/HouZLYWLLLGW24} and LLM-based SE agents~\cite{DBLP:journals/corr/abs-2409-02977} span a wide spectrum of development tasks without treating migration as a dedicated, first-class engineering problem. 
In related niches, a survey of GUI agents focuses on operating existing graphical interfaces rather than migrating them~\cite{DBLP:journals/tmlr/ZhangHQ0LQ0MLLR25}, a tertiary study on mobile software engineering identifies limited secondary-study coverage of evolution, maintenance, and cross-platform development~\cite{DBLP:journals/infsof/ZeinSG23}, and classic literature on legacy system migration predates the advent of LLMs~\cite{DBLP:journals/software/BisbalLWG99}. 
Crucially, none of these surveys traces the progression from module-level transformation to repository-level integration, nor do they systematically link LLM-assisted techniques to the empirical evidence used to validate and accept their results.

To bridge this critical gap, this survey systematically analyzes 96 sources: 46 primary studies on direct migration, 30 studies providing adjacent technical or evaluation support, and 20 foundational background sources (Section~\ref{sec:method}). 
We organize our analysis around four complementary perspectives:
\begin{itemize}
  \item \textbf{Software Engineering (SE) perspective:} examines which lifecycle activities LLMs support, spanning requirements engineering, architectural design, software construction, and testing.
  \item \textbf{Role-based perspective:} characterizes whether models act as direct \emph{editors} that generate target code or as \emph{orchestrators} that operate external tools to build, verify, and test the software.
  \item \textbf{Agent perspective:} investigates workflow control mechanisms, contrasting deterministic (agentless) pipelines, single-agent loops, and multi-agent coordination.
  \item \textbf{Technique perspective:} explores the foundational techniques these systems rely on, namely knowledge engineering, project context retrieval, environment feedback loops, and model adaptation.
\end{itemize}
Building upon these perspectives, we systematically compare existing evaluation practices and benchmark suites to address five concrete research questions:
\begin{itemize}
  \item \textbf{RQ1:} Which migration tasks, target domains, and granularities do existing studies cover?
  \item \textbf{RQ2:} How do LLMs support distinct engineering activities throughout the migration lifecycle?
  \item \textbf{RQ3:} What functional roles do models play, and how are their workflows orchestrated?
  \item \textbf{RQ4:} Which enabling techniques empower LLM-assisted migration?
  \item \textbf{RQ5:} What verification properties do current evaluations establish, and what fundamental challenges remain open?
\end{itemize}

Guided by these questions, our analysis yields four primary findings. 
First, although repository-level tasks account for 30 of the 46 direct migration studies, acceptance evidence overwhelmingly relies on build success, supplied tests, or static and model-based proxies, leaving verified runtime execution on target platforms remarkably rare. 
Second, requirements engineering remains the least supported lifecycle activity among the examined systems, with most approaches implicitly treating legacy source code and pre-existing tests as the \emph{de facto} specification. 
Third, most multi-agent architectures are simple linear pipelines of stage-specific agents with rigid hand-offs, yet few studies benchmark them against single-agent baselines under strictly matched computational budgets. 
Fourth, compilation, visual similarity, test outcomes, and runtime application behavior evaluate complementary properties with scope-bounded evidence; consequently, reported results cannot be meaningfully conflated into a single, monolithic success rate.

In summary, this survey makes the following contributions:
\begin{itemize}
  \item A task and evidence framework that separates direct migration studies
        from adjacent support and background material, and codes migration
        granularity and evidence status independently
        (Section~\ref{sec:method}).
  \item An analysis of LLM-assisted migration from four
        complementary perspectives: SE activities, model roles, orchestration
        structures, and enabling techniques
        (Sections~\ref{sec:se}--\ref{sec:technique}).
  \item A comparison of evaluation practices and benchmark resources that
        clarifies what current results can and cannot show about
        repository-level migration (Section~\ref{sec:eval}).
  \item A research agenda that turns the identified gaps into testable
        questions with proposed data, baselines, and outcome measures
        (Section~\ref{sec:challenges}).
\end{itemize}

\input{figures/Taxonomy.tex}

Figure~\ref{fig:structure} summarizes the structure of this survey.
Section~\ref{sec:background} introduces background on software migration and
LLM-based agents, and Section~\ref{sec:method} describes the review
methodology. Sections~\ref{sec:se}--\ref{sec:technique} present the analyses
from the SE, role-based, agent, and technique perspectives.
Section~\ref{sec:eval} compares evaluation practices and benchmark resources,
Section~\ref{sec:challenges} discusses challenges and open problems, and
Section~\ref{sec:conclusion} concludes.

%% file: figures/Taxonomy.tex
\begin{figure*}[!htbp]
  \centering 
  \begin{forest}
  for tree={
    font=\small, 
    rounded corners=2pt,
    child anchor=west,
    parent anchor=east,
    grow'=east,  
    draw=darkblue,
    anchor=west,
    node options={align=center},
    inner ysep=3pt,
    s sep=3pt,
    l sep=12pt,
    edge path={
      \noexpand\path[\forestoption{edge}]
        (.child anchor) -| +(-5pt,0) -- +(-5pt,0) |-
        (!u.parent anchor)\forestoption{edge label};
    },
    where level=0{text width=3.5cm}{}, 
    where level=1{text width=5.2cm}{}, 
    where level=2{text width=5.5cm}{}  
  },
  [LLM-Assisted Software Migration
    [Background \S\ref{sec:background}]
    [Survey Methodology \S\ref{sec:methodology}]
    [SE Perspectives \S\ref{sec:se}
        [Requirements Engineering \S\ref{sec:se-spec}]
        [Software Design \S\ref{sec:se-design}]
        [Software Construction \S\ref{sec:se-construction}]
        [Testing \S\ref{sec:se-testing}]
        [End-to-end Software Development \S\ref{sec:se-dev}]
        [End-to-end Software Maintenance \S\ref{sec:se-maint}]
    ]
    [Role-Based Perspectives \S\ref{sec:role}
        [LLM-as-Editor \S\ref{sec:se-editor}]
        [LLM-as-Orchestrator \S\ref{sec:sub-llm-orchestrator}]
    ]
    [Agent Perspectives \S\ref{sec:agent}
        [Agentless Workflows \S\ref{sec:orch-agentless}]
        [Single-Agent Migration \S\ref{sec:orch-single}]
        [Multi-Agent Migration \S\ref{sec:orch-multi}]
    ]
    [Enabling Techniques \S\ref{sec:technique}
        [Knowledge Engineering \S\ref{subsec:knowledge-engineering}]
        [Project Context \S\ref{subsec:project-context}]
        [Environment \& Feedback \S\ref{subsec:environment-feedback}]
        [Model Adaptation \S\ref{subsec:model-adaptation}]
    ]
    [Evaluation Practices and Benchmark Resources \S\ref{sec:eval}]
    [Challenges and Open Problems \S\ref{sec:challenges}]
  ]
  \end{forest}
  \caption{\small Structure of This Survey.} 
  \label{fig:structure}
\end{figure*}
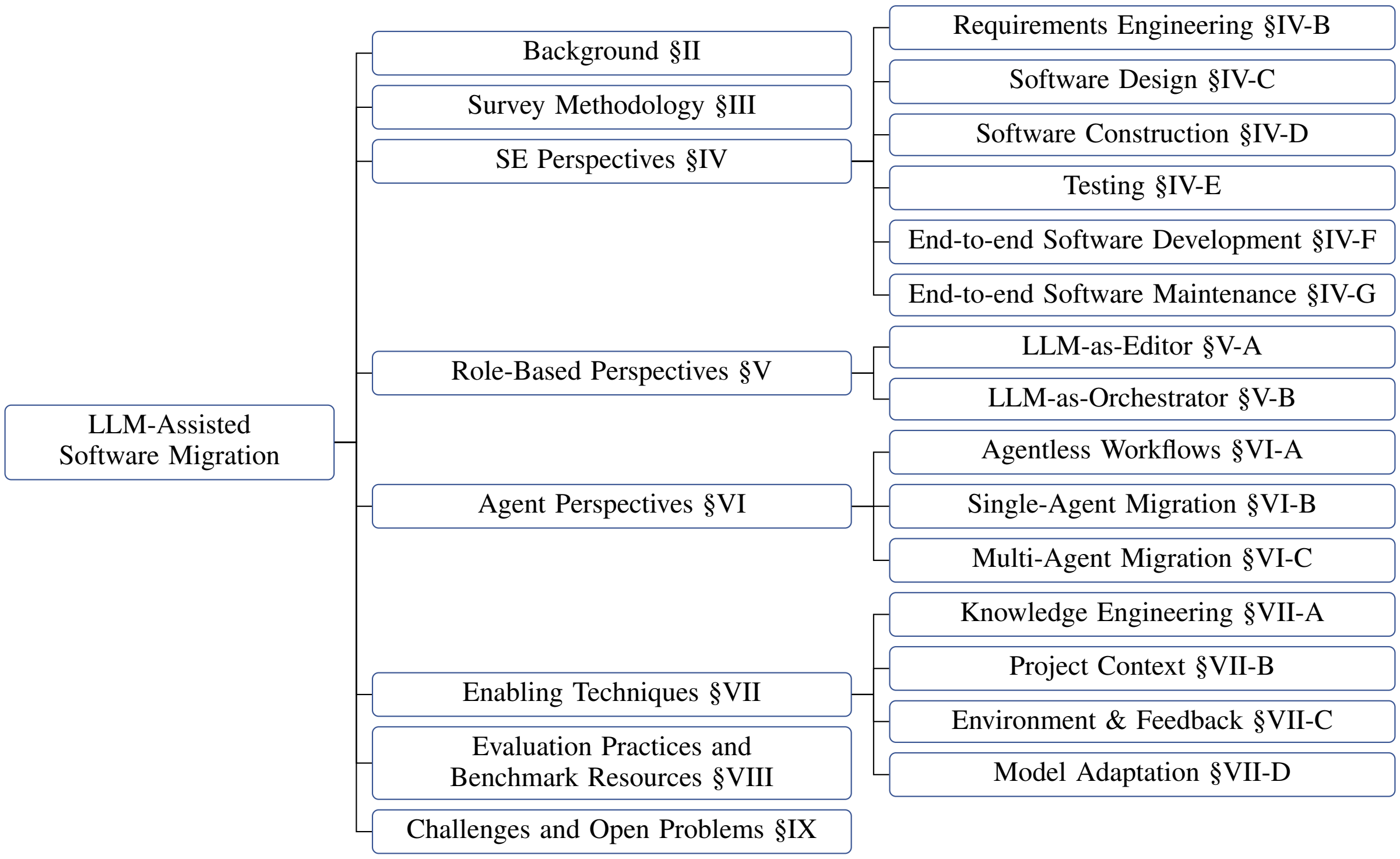

%% file: sections/background.tex
\section{Background and Preliminaries}
\label{sec:background}

This section introduces the foundational concepts upon which the remainder of this survey is built.
We first define software migration and its associated terminology
(Section~\ref{sec:bg-migration}), followed by a classification of migration based on the nature of the changes
and the scope of the migrated unit (Section~\ref{sec:bg-types}). Next, we
summarize the core concepts of LLMs and agents utilized in subsequent sections
(Section~\ref{sec:bg-llm}), and finally, we position our work in the context of related
surveys (Section~\ref{sec:bg-related}).

\subsection{Software Migration}
\label{sec:bg-migration}

\paragraph{Task formulation}
A migration task utilizes an existing implementation, available information, a target environment, and an acceptance contract to produce a target implementation, alongside evidence that the contract holds. The available information may include source code, documentation, screenshots, runtime traces, tests, and target scaffolds, though few tasks supply all. The target environment defines the required language, framework, platform, and tool versions. The acceptance contract specifies which user-observable behaviors must be preserved and which platform differences are acceptable. Pixel-perfect replication or an identical internal architecture is rarely required. Buildability, executability, and behavior preservation are distinct requirements; Section~\ref{sec:eval} details how each is evaluated.

\paragraph{Related terms}
The literature often uses \emph{translation}, \emph{migration}, and
\emph{modernization} loosely; indeed, many studies that migrate whole
repositories describe their task simply as code
translation~\cite{DBLP:journals/pacmse/IbrahimzadaKPAPSJ25, DBLP:journals/tse/WangWWGCGLMMZZ26}.
We therefore distinguish these terms by what they change and what they must
preserve. \emph{Translation} alters the representation of a code unit,
typically its language, and succeeds when the unit retains its original semantics.
\emph{Migration} additionally changes the environment in which the code runs, requiring the adaptation of
dependencies, APIs, and platform contracts; it succeeds only when
the system's user-observable behavior is preserved within that new environment.
\emph{Refactoring}, in contrast, preserves external behavior while
changing the internal
structure~\cite{DBLP:journals/tse/MensT04}, and in this survey we use the term only where the environment also stays the same, whereas
\emph{modernization} aims to improve architecture or quality attributes and
need not preserve the original
design~\cite{comella2000survey}. All four are
forms of reengineering, which involves examining an existing system and subsequently
altering it~\cite{DBLP:journals/software/ChikofskyC90}. The
horseshoe model illustrates where these processes
diverge~\cite{DBLP:conf/wcre/KazmanWC98}:
translation can remain strictly at the code level, whereas migration often requires
recovering higher-level artifacts, such as specifications and architecture,
before rebuilding the system for the target. Throughout this survey, we
classify a study based on the task it evaluates rather than the label it uses; thus, a
``translation'' study counts as migration if its result is evaluated in the
target environment.

\subsection{Migration Types and Granularity}
\label{sec:bg-types}

\paragraph{Migration types}
Table~\ref{tab:migration-types} categorizes migration based on the elements that change between
the source and the target. Cross-language migration involves changing the programming
language; version and API migration updates the codebase to accommodate new language releases,
libraries, or SDKs; framework migration substitutes component and state
abstractions while remaining within the same language; and platform migration transitions
an application to a different operating system or deployment platform. Test migration is an adjacent task
that focuses on transferring test assets rather than the implementation code itself. These types frequently
co-occur: for instance, porting an Android application to HarmonyOS simultaneously entails changes to the
language, the UI framework, and the underlying platform.

\begin{table*}[t]
\caption{Migration types and representative studies. The provided examples illustrate the task
scope rather than equivalent evidence of successful delivery.}
\label{tab:migration-types}
\centering\small
\begin{tabular}{p{0.14\textwidth}p{0.26\textwidth}p{0.52\textwidth}}
\toprule
Type & Typical changes & Representative studies \\
\midrule
Cross-language & Syntax, types, memory model, standard library, idioms &
  Repository-level translation~\cite{DBLP:journals/pacmse/IbrahimzadaKPAPSJ25, DBLP:journals/corr/abs-2604-07341}; C-to-Rust project
  migration~\cite{DBLP:conf/iceccs/CaiLHYWLWYJ25} \\
Version and API & Deprecated APIs, breaking dependency updates, language or
  SDK versions &
  Java version upgrades~\cite{DBLP:conf/kdd/LiuLZCLNOSHTD26}; Android deprecated API
  updates~\cite{DBLP:journals/tse/MahmudDCYNY26}; breaking dependency
  updates~\cite{DBLP:journals/ese/ReyesMBNBM26} \\
Framework & Component, state, and plugin abstractions; build system &
  Enterprise Java framework migration~\cite{DBLP:journals/corr/abs-2605-06754}; Android
  XML-to-Compose UI conversion~\cite{DBLP:journals/pacmse/GaoHYX26} \\
Platform & Language, UI framework, runtime calls, resources, permissions,
  configuration &
  Android-to-HarmonyOS repository migration~\cite{DBLP:journals/corr/abs-2608-09273};
  Android-to-iOS translation~\cite{DBLP:journals/corr/abs-2507-16037}; UI translation to
  ArkUI~\cite{DBLP:conf/internetware/GongWCHW25} \\
Test assets & Actions, locators, and assertions of existing tests &
  GUI test transfer~\cite{DBLP:conf/issta/QinZW19}; Android-to-OpenHarmony test
  migration~\cite{DBLP:journals/corr/abs-2607-11245} \\
\bottomrule
\end{tabular}
\end{table*}

\paragraph{Granularity}
Migration tasks also vary in the scope of the migrated unit. This dimension
is largely orthogonal to the migration type, although platform migration
typically involves larger units. Table~\ref{tab:granularity} delineates two
levels based on their required deliverables. \emph{Module-level} migration transforms
specific units, such as functions, files, API usages, or UI pages; the remainder of
the codebase may be read as context but remains unaltered.
\emph{Repository-level} migration delivers a modified codebase whose components must
operate cohesively; thus, interfaces, dependencies, tests, and build configurations
must remain consistent across all files. Some studies call this level
\emph{project-level}; we treat the two terms as synonyms and define the level
by the deliverable, not by the boundary of a version-control repository.

The repository level includes the delivery of complete applications. An
application is delivered as a repository; what sets it apart is a further
requirement, namely that the migrated code be deployed and run on the target
platform, interoperating with resources, configurations, and platform services
and exhibiting the expected user-visible behavior. We record this
\emph{deployed-application requirement} as a property of a repository-level
task and not as a third level, because it concerns how the result must be
shown to work rather than how much code is migrated, and because only four
studies in our corpus carry it (Section~\ref{sec:collection}). A task carries
the requirement when its task statement, workflow, or evaluation requires the
migrated repository to be deployed and launched on the target platform. Whether
deployment actually took place, whether runtime probes were used, and whether
behavior was accepted by checks independent of the system's own feedback are
recorded separately. We still use ``application-level'' informally for properties of a running application, such as application-level acceptance, but not as a granularity level. Each step subsumes the
requirements of the preceding one while introducing its own. Consequently, success at one level
does not guarantee success at the next.

We decide the level by the unit of output that the evaluated task asks the
system to produce, not by the stated long-term aim of a study, the scope of
the context a method reads, or the unit at which the result was validated.
A type-migration method motivated by repository-level translation that
produces and evaluates variable types is
module-level~\cite{DBLP:conf/soap-ws/XuH25}; RepoTransAgent translates one
function at a time with repository context and is module-level~\cite{DBLP:journals/corr/abs-2508-17720}; and
the Android-to-iOS pilot study produces a whole app project but validates
individual files and is repository-level~\cite{DBLP:journals/corr/abs-2507-16037}.

\begin{table}[t]
\caption{Granularity levels of software migration. The deployed-application requirement applies to some repository-level tasks.}
\label{tab:granularity}
\centering\small
\setlength{\tabcolsep}{4pt}
\begin{tabular}{p{0.19\linewidth}p{0.37\linewidth}p{0.32\linewidth}}
\toprule
Level & Additional requirements & Typical acceptance evidence \\
\midrule
Module & Correct translation of the selected unit & Compilation, unit
  tests, similarity metrics \\
Repository & Dependency ordering, interface consistency, coordinated edits
  to code, tests, and build files & Project build, test suites, coverage \\
\hspace{0.5em}+ deployed & As a running application: resources, configuration, platform services, user-visible
  behavior & Deployment, GUI-level tests, behavioral checks \\
\bottomrule
\end{tabular}
\end{table}

\subsection{LLMs and LLM-Based Agents for Code}
\label{sec:bg-llm}

\paragraph{Code LLMs}
Large language models (LLMs) pretrained on extensive corpora of code and natural language possess the capability to generate, translate, and repair code based on instructions and
examples~\cite{DBLP:journals/tmlr/LiAZMKMMALCLZZW23, DBLP:journals/corr/abs-2308-12950, DBLP:journals/corr/abs-2402-19173}. These models are typically leveraged via prompting techniques---such as zero-shot and
few-shot prompting, retrieval-augmented prompting, and iterative prompting
with execution feedback---or they can be tailored to specific tasks through
fine-tuning~\cite{DBLP:journals/tosem/HouZLYWLLLGW24}. Two primary limitations are particularly relevant to software migration: first, the
context window restricts the proportion of a project the model can process simultaneously; second, the
model may generate code that appears plausible but is factually incorrect, such as invoking APIs that do
not exist in the target environment.

\paragraph{LLM-based agents}
An LLM-based agent employs an LLM as its central controller, augmenting it with capabilities for
planning, memory, perception, and action execution (e.g., invoking compilers, test
runners, or search utilities)~\cite{DBLP:journals/corr/abs-2409-02977}. Such agents can operate autonomously or collaboratively within multi-agent
systems, where agents assuming distinct roles exchange intermediate results.
It is important to note that not every LLM-based system qualifies as an agent; numerous migration systems simply invoke an LLM at
predetermined stages of a pipeline governed by hard-coded control flows. We will revisit
this distinction in Section~\ref{sec:agent} and further explore the specific roles a model
can assume in Section~\ref{sec:role}.

\subsection{Related Surveys}
\label{sec:bg-related}

Several existing surveys explore topics adjacent to the focus of this paper.

\emph{Code translation.} Chen et al.\ review neural code translation,
comprehensively covering task characteristics, code modeling, model construction,
post-processing, and evaluation metrics~\cite{DBLP:journals/corr/abs-2505-07425}. Their review also covers file- and project-level translation, including repository dependencies and project-level tests; however, it organizes the work by translation technique and metric rather than by the deliverable a migration must produce, the control over the process, or the acceptance evidence.
Additionally, Zhang et al.\ survey the application of LLMs within compilers, which includes
transpilation techniques~\cite{DBLP:journals/ccfthpc/ZhangZYXWFC26}. 

\emph{LLMs and agents for software engineering.} Hou et al.\ review the application of LLMs
across a broad spectrum of software engineering tasks~\cite{DBLP:journals/tosem/HouZLYWLLLGW24}, while Liu et al.\ analyze
LLM-based agents in software engineering from both software engineering and agent-centric
perspectives~\cite{DBLP:journals/corr/abs-2409-02977}. Neither survey treats migration as a task of its own: the relevant work is spread across other categories such as code translation, without delving into
its unique requirements.

\emph{Mobile software engineering and GUI agents.} Zhang et al.\ survey
LLM-based GUI agents, which are designed to operate existing applications rather than migrate
them~\cite{DBLP:journals/tmlr/ZhangHQ0LQ0MLLR25}. Zein et al.\ map secondary studies in
mobile application software engineering, identifying limited secondary-study coverage of software evolution, maintenance, and cross-platform development~\cite{DBLP:journals/infsof/ZeinSG23}.

\emph{Legacy migration and modernization.} Earlier surveys have investigated the
migration and modernization of legacy systems, though these studies predate the advent of
LLMs~\cite{DBLP:journals/software/BisbalLWG99, comella2000survey}.

This survey distinguishes itself from the aforementioned works in three key respects. First, it traces migration from module-level transformations to repository-level integration across various
migration types. Second, it provides an in-depth analysis of how LLM-assisted systems are architected, ranging from the specific roles the models play to the underlying techniques that support them. Finally, it systematically correlates the methodologies of each reviewed study with the acceptance evidence they report.

%% file: sections/methodology.tex
\section{Review Methodology}
\label{sec:method}
\label{sec:methodology}

This survey is designed as a structured, research-question-driven review rather than an
exhaustive systematic literature review. We draw on established guidelines for
systematic mapping~\cite{DBLP:conf/ease/PetersenFMM08} and
snowballing~\cite{DBLP:conf/ease/Wohlin14} within software engineering. This section details how we collected and selected the
literature (Sections~\ref{sec:search} and~\ref{sec:selection}), summarizes the
resulting corpus (Section~\ref{sec:collection}), and establishes the analytical
framework applied throughout the remainder of the survey (Section~\ref{sec:coding}).

\subsection{Literature Search}
\label{sec:search}

We searched arXiv, OpenReview, Google Scholar, and IEEE Xplore. Our queries combined three groups of
terms: model terms (\emph{large language model}, \emph{LLM}, \emph{agent}),
task terms (\emph{migration}, \emph{code translation}, \emph{transpilation},
\emph{porting}, \emph{upgrade}, \emph{modernization}), and target terms
(\emph{repository}, \emph{application}, \emph{API}, \emph{framework},
\emph{version}, \emph{Android}, \emph{HarmonyOS}, \emph{iOS}). We collected papers published from January 2018 to October 2026; the last
search was conducted on October 10, 2026. Earlier work is retained only as
background on terminology and reengineering or as non-LLM comparison work. To find relevant studies that
the keyword search missed, we then applied backward and forward snowballing
to the selected primary
studies~\cite{DBLP:conf/ease/Wohlin14}.

\subsection{Selection Criteria}
\label{sec:selection}

Table~\ref{tab:eligibility} details the selection criteria. Each accepted
record enters the corpus under one of three classes. Class~A comprises studies
that use an LLM to directly migrate code across languages, versions, APIs, frameworks, or
platforms at any level of granularity. Class~B encompasses migration methods that do not use an LLM and studies of adjacent tasks,
such as testing, repair, context recovery, and GUI operation, whose underlying mechanisms
or evaluation protocols can inform migration. Class~C provides foundational background on
terminology, prior surveys, review methodology, and code models. We treat a
preprint and its subsequent formal publication as a single study, reporting
results exclusively from the version we cite. Studies for which the full text was
inaccessible were excluded. Function- and file-level translation evaluated in isolation is reviewed elsewhere~\cite{DBLP:journals/corr/abs-2505-07425}; we include module-level studies only selectively, as reference points for the repository-level work, so the repository-level studies form our main sample and every proportion by granularity describes our corpus rather than the field.

\begin{table}[t]
\caption{Selection criteria.}
\label{tab:eligibility}
\centering\small
\begin{tabular}{p{0.22\columnwidth}p{0.68\columnwidth}}
\toprule
Class & Criterion \\
\midrule
A: direct migration & Uses an LLM to migrate implementation code, or evaluates LLMs or LLM-based agents on an explicit
  migration task, at the module or repository level. \\
B: adjacent support & Rule-based and historical migration methods kept for comparison; studies of testing, repair, context recovery, or GUI
  operation with mechanisms or evaluation protocols applicable to migration;
  not counted as implementation migration. \\
C: background & Terminology, prior surveys, review methodology, and
  foundational code models; not counted as evaluated migration
  methods. \\
Excluded & Transfer of user data without implementation changes; translation
  tasks unrelated to code; full text unavailable. \\
\bottomrule
\end{tabular}
\end{table}

\subsection{Collection Results}
\label{sec:collection}

The final corpus comprises 96 records: 46 in Class~A, 30 in Class~B, and 20 in
Class~C (Table~\ref{tab:corpus-roles}). Of the 46 direct migration studies,
30 (65.2\%) address repository-level tasks and 16 (34.8\%) target module-level artifacts such as
APIs, UI pages, functions, and types (Table~\ref{tab:classA-distribution}). Four of the repository-level studies carry the
deployed-application requirement, and they differ in how far it was met (Table~\ref{tab:deployed}). By primary migration type, 23 studies are
cross-language (14 of them C-to-Rust), 17 concern versions and APIs, 5 change the platform, and 1 changes the framework. Five Class~A
studies specifically concern mobile applications, and 24 of the 46 are preprints in the version we cite. It should be noted that granularity here reflects the unit of output that the task requires (Section~\ref{sec:bg-types}),
rather than the level at which the results were successfully validated;
we analyze this acceptance evidence thoroughly in Section~\ref{sec:eval}.

\input{tables/corpus_composition}

\input{tables/classA_distribution}

\begin{table}[t]
\caption{The four repository-level studies that carry the deployed-application
requirement, and how far it was met. NR: not reported in the cited account.}
\label{tab:deployed}
\centering\small
\setlength{\tabcolsep}{3pt}
\begin{tabular}{p{0.24\columnwidth}p{0.20\columnwidth}p{0.15\columnwidth}p{0.29\columnwidth}}
\toprule
Study & Deployment performed & Runtime probes & Independent behavioral acceptance \\
\midrule
ScarfBench~\cite{DBLP:journals/corr/abs-2605-06754} & Yes (container) & NR & Yes \\
Mono2Sls~\cite{DBLP:journals/corr/abs-2604-24550} & Yes (cloud) & NR & Yes, after the workflow \\
ECAT~\cite{DBLP:journals/corr/abs-2608-09273} & Yes (emulator) & Yes & No (model-judged) \\
Pilot study~\cite{DBLP:journals/corr/abs-2507-16037} & No & No & No \\
\bottomrule
\end{tabular}
\end{table}

\subsection{Analytical Framework and Coding}
\label{sec:coding}

We analyze the Class~A and Class~B studies from four distinct perspectives, culminating
in an analysis of their evaluation methodologies (Table~\ref{tab:framework}). The \emph{SE
perspective} examines the specific migration activities a study supports, ranging
from specification and design to construction and testing. The
\emph{role-based perspective} investigates whether the model is tasked with writing target code or
operating peripheral tools. The \emph{orchestration perspective} determines the locus of control for
step-by-step decision-making: hard-coded program logic, a single agent, or a multi-agent system. The
\emph{technique perspective} explores the underlying techniques used to equip the model
with necessary information and capabilities. A single study is coded separately on all four dimensions, although the role and orchestration perspectives are linked by definition: a model can direct the process only where the orchestration structure gives it control over the next step. The final column of Table~\ref{tab:framework} details the
information extracted for each perspective. For every study, we also record
its canonical citation, cited version, publication status, and assigned class.
Where a paper does not report enough detail to decide a classification, we follow the authors' description and mark the entry as self-described. Activities (Table~\ref{tab:stage-artifacts}) and roles and orchestration (Table~\ref{tab:agent-architectures}) are coded for the systems discussed in detail, not for all 46 studies, so statements about them are qualitative observations about those systems.

\begin{table*}[t]
\caption{Analytical framework: research questions, perspectives, categories, and
extracted information.}
\label{tab:framework}
\centering\small
\begin{tabular}{p{0.05\textwidth}p{0.15\textwidth}p{0.30\textwidth}p{0.40\textwidth}}
\toprule
RQ & Perspective & Categories & Extracted information \\
\midrule
RQ1 & Task landscape (\S\ref{sec:collection}) & Migration type; module,
  repository & Source and target languages, frameworks,
  platforms, and versions; inputs and outputs; granularity \\
RQ2 & SE (\S\ref{sec:se}) & Specification, design, construction, testing &
  Specification recovery; decomposition and mapping; generation and
  generate--check--repair loops; acceptance checks; human participation \\
RQ3 & Role (\S\ref{sec:role}) & Writing target code; operating tools around
  code & Model and version; artifacts the model produces; tools it invokes;
  division of work with deterministic components \\
RQ3 & Orchestration (\S\ref{sec:agent}) & Agentless, single-agent,
  multi-agent & Who chooses the next step; stopping rule and retry budget;
  agent topology \\
RQ4 & Technique (\S\ref{sec:technique}) & Knowledge engineering, project
  context engineering, environment and feedback engineering, model
  adaptation & Migration knowledge and mappings; retrieved project context;
  environment and feedback signals; fine-tuning data and objectives \\
RQ5 & Evaluation (\S\ref{sec:eval}) & Property checked, scope, means,
  judge & Benchmark; sample unit and denominator;
  metrics and oracle; baselines; cost \\
\bottomrule
\end{tabular}
\end{table*}

\subsection{Threats to Validity}
\label{sec:threats}

Four properties of the available evidence affect how our results should be
read. First, many of the included studies are recent preprints whose results
may change between versions. We therefore record, for every extracted result,
the version from which it is taken, so that later changes can be traced.
Second, the gaps identified in this survey describe the published evidence
available to us, not the absence of the corresponding work in practice;
unpublished or ongoing work may already address some of them. The open
problems in Section~\ref{sec:challenges} should accordingly be read as gaps
in the published literature rather than as claims about industrial practice.
Third, the review is not exhaustive and covers module-level translation only selectively, so statements that we found no study of some kind mean that none appeared in this corpus.
Fourth, assigning a class, a primary type, and a granularity involves judgment.

%% file: tables/corpus_composition.tex
\begin{table*}[t]
\caption{Corpus composition. Every cited source appears in exactly one row group. Class~A studies are grouped by the granularity of the deliverable their task targets; the repository level includes tasks that must deliver a deployed application.}
\label{tab:corpus-roles}
\centering\scriptsize
\begin{tabular}{p{0.24\textwidth}p{0.04\textwidth}p{0.64\textwidth}}
\toprule
Class and group & Count & Records \\
\midrule
\textbf{A: LLM-assisted direct migration} & \textbf{46} & \\
\quad Module & 16 & \cite{DBLP:journals/corr/abs-2607-04092, DBLP:journals/tse/MahmudDCYNY26, DBLP:conf/iclr/LeeMKC0CWR24,  zheng2026portingdeclarativeuiharmonyos, DBLP:conf/internetware/GongWCHW25, DBLP:journals/corr/abs-2609-03592, DBLP:journals/corr/abs-2602-09930, DBLP:conf/soap-ws/XuH25, DBLP:journals/corr/abs-2605-21537, DBLP:journals/corr/abs-2503-23791, DBLP:journals/corr/abs-2607-19966, DBLP:conf/icse/Macedo00CA25, DBLP:journals/corr/abs-2505-04852, DBLP:conf/kbse/RamosMLMMG23, DBLP:journals/corr/abs-2602-01107, DBLP:journals/corr/abs-2508-17720} \\
\quad Repository & 30 & \cite{DBLP:conf/iceccs/CaiLHYWLWYJ25, DBLP:conf/kdd/LiuLZCLNOSHTD26, DBLP:journals/ese/ReyesMBNBM26, DBLP:journals/corr/abs-2608-09273, DBLP:journals/corr/abs-2507-16037, DBLP:conf/sigsoft/ZiftciNSKCK25, DBLP:journals/tse/WangWWGCGLMMZZ26, DBLP:journals/pacmse/IbrahimzadaKPAPSJ25, DBLP:journals/corr/abs-2604-07341, DBLP:journals/corr/abs-2605-06754, DBLP:journals/pacmse/BairiSKCIPRAS24, DBLP:conf/eacl/FujiiMYS26, DBLP:journals/corr/abs-2510-04852, DBLP:journals/corr/abs-2605-14634, DBLP:journals/corr/abs-2603-28686, DBLP:journals/pacmse/YuanMCSWLP26, DBLP:journals/ese/HongR25, DBLP:conf/ijcai/KuangXYYWKX25, DBLP:journals/corr/abs-2608-14128, DBLP:journals/pacmpl/ZhangDWPK25, DBLP:conf/icse-seip/WangYSWCZSXG26, wang2026his2transknowledgeguidedagenticframework, DBLP:journals/corr/abs-2604-24550, DBLP:journals/tse/NitinKVR26, DBLP:journals/corr/abs-2606-24446, DBLP:conf/esem/AlmeidaXV24, DBLP:conf/agent/AlmeidaVX26, DBLP:journals/corr/abs-2504-13272, DBLP:journals/corr/abs-2604-12048, DBLP:journals/corr/abs-2505-15858} \\
\quad \emph{of which require a deployed application} & 4 & \cite{DBLP:journals/corr/abs-2608-09273, DBLP:journals/corr/abs-2507-16037, DBLP:journals/corr/abs-2605-06754, DBLP:journals/corr/abs-2604-24550} \\
\quad \emph{of which mobile} & 5 & \cite{DBLP:journals/tse/MahmudDCYNY26, DBLP:journals/corr/abs-2608-09273, DBLP:journals/corr/abs-2507-16037, zheng2026portingdeclarativeuiharmonyos, DBLP:conf/internetware/GongWCHW25} \\
\midrule
\textbf{B: adjacent and comparison work} & \textbf{30} & \\
\quad Non-LLM migration method & 5 & \cite{DBLP:journals/pacmpl/EmreSDH21, DBLP:journals/pacmse/GaoHYX26, DBLP:journals/mis/LamhaddabLE19, DBLP:journals/monet/PuderA13, DBLP:conf/nips/RoziereLCL20} \\
\quad Test migration & 3 & \cite{DBLP:journals/corr/abs-2607-11245, DBLP:conf/issta/QinZW19, DBLP:journals/pacmse/BeyzaeiTRMM25} \\
\quad Repair benchmark or agent environment & 3 & \cite{DBLP:journals/corr/abs-2608-18588, xie2026arkevalbenchmarkingevaluatingautomated, DBLP:conf/iclr/RawlesCCWLFLBLC25} \\
\quad General code benchmark & 6 & \cite{DBLP:conf/iclr/0003XM24, DBLP:conf/nips/LiuXW023, DBLP:journals/corr/abs-2107-03374, DBLP:journals/tse/CassanoGNNPPYZAFGGJ23, DBLP:conf/iclr/JimenezYWYPPN24, DBLP:conf/iclr/JainHGLYZWSSS25} \\
\quad General LLM technique & 5 & \cite{DBLP:conf/iclr/ChenLSZ24, DBLP:conf/nips/MadaanTGHGW0DPY23, wu2023autogenenablingnextgenllm, DBLP:conf/iclr/HongZCZCWZWYLZR24, DBLP:conf/acl/QianLLCDL0CSCXL24} \\
\quad Mobile platform study & 5 & \cite{DBLP:conf/esem/JoorabchiMK13, DBLP:conf/issta/RiganelliMRMM20, DBLP:conf/icsm/McDonnellRK13, DBLP:conf/kbse/WeiLC16, DBLP:journals/ese/ScalabrinoBLPLO20} \\
\quad Practice and productivity evidence & 2 & \cite{chand_et_al:LIPIcs.ESEM.2026.85, DBLP:journals/corr/abs-2507-09089} \\
\quad GUI model recovery & 1 & \cite{DBLP:conf/wcre/MemonBN03} \\
\midrule
\textbf{C: background} & \textbf{20} & \\
\quad Prior survey & 8 & \cite{DBLP:journals/corr/abs-2505-07425, DBLP:journals/tosem/HouZLYWLLLGW24, DBLP:journals/corr/abs-2409-02977, DBLP:journals/tmlr/ZhangHQ0LQ0MLLR25, DBLP:journals/infsof/ZeinSG23, DBLP:journals/software/BisbalLWG99, comella2000survey, DBLP:journals/ccfthpc/ZhangZYXWFC26} \\
\quad Terminology and reengineering & 3 & \cite{DBLP:journals/tse/MensT04, DBLP:journals/software/ChikofskyC90, DBLP:conf/wcre/KazmanWC98} \\
\quad Review methodology & 3 & \cite{DBLP:conf/ease/PetersenFMM08, DBLP:conf/ease/Wohlin14, DBLP:journals/infsof/KitchenhamBBTBL09} \\
\quad Code model & 6 & \cite{DBLP:journals/tmlr/LiAZMKMMALCLZZW23, DBLP:journals/corr/abs-2308-12950, DBLP:journals/corr/abs-2402-19173, DBLP:conf/emnlp/0034WJH21, DBLP:conf/naacl/AhmadCRC21, DBLP:conf/iclr/NijkampPHTWZSX23} \\
\bottomrule
\end{tabular}
\end{table*}

%% file: tables/classA_distribution.tex
\begin{table}[t]
\caption{Distribution of the 46 Class~A studies by primary migration type and target granularity, with contribution kind and publication status.}
\label{tab:classA-distribution}
\centering\small
\begin{tabular}{lrrr}
\toprule
Primary type & Module & Repository & Total \\
\midrule
Cross-language & 8 & 15 & 23 \\
Version and API & 6 & 11 & 17 \\
Framework & 0 & 1 & 1 \\
Platform & 2 & 3 & 5 \\
\midrule
Total & 16 & 30 & 46 \\
Share & 34.8\% & 65.2\% & \\
\midrule
\multicolumn{4}{p{0.92\columnwidth}}{\emph{Deployed application required:} 4 of the 30 repository-level studies.} \\
\multicolumn{4}{p{0.92\columnwidth}}{\emph{Contribution kind:} 33 methods or systems, 7 empirical or experience studies, 6 benchmarks.} \\
\multicolumn{4}{p{0.92\columnwidth}}{\emph{Publication status of the cited version:} 22 peer-reviewed, 24 preprints.} \\
\multicolumn{4}{p{0.92\columnwidth}}{\emph{Year of the cited version:} 2023: 1, 2024: 3, 2025: 16, 2026: 26.} \\
\bottomrule
\end{tabular}
\end{table}

%% file: sections/se_perspective.tex
\section{SE Perspective: Migration Activities}
\label{sec:se}

Adopting a software engineering perspective to address RQ2, this section investigates the specific lifecycle activities that LLM-assisted migration systems support, alongside the artifacts each activity produces. We first establish our analytical framework (Section~\ref{sec:se-overview}). Subsequently, Sections~\ref{sec:se-spec}--\ref{sec:se-testing} delve into the four foundational migration activities: requirements engineering, software design, software construction, and testing. Finally, Sections~\ref{sec:se-dev} and~\ref{sec:se-maint} explore how existing systems synthesize these discrete activities into cohesive, end-to-end migration workflows.

\subsection{Overview of Migration Activities}
\label{sec:se-overview}

\begin{figure*}[t]
  \centering
  \includegraphics[width=\textwidth]{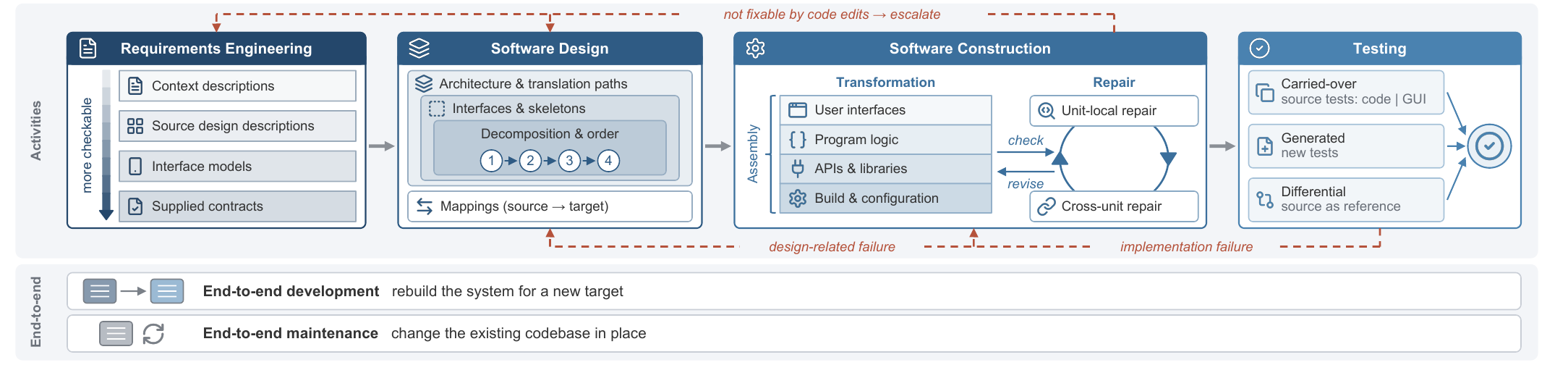}
  \caption{Migration activities analyzed in this section. Each panel lists the
categories discussed in the corresponding subsection. Gray arrows show the
usual flow of artifacts between activities, and the blue loop shows iterative
refinement between transformation and repair within construction. Red dashed
arrows show feedback to earlier activities: implementation failures found by
testing return to repair, design-related failures return to design, and
failures that code edits cannot fix return to requirements engineering or
software design. The two
bands denote end-to-end systems that span all activities.}
  \label{fig:se-activities}
\end{figure*}

Migration, much like traditional software development, involves determining what the target system must achieve, how it should be structured, how it is constructed, and whether the outcome is acceptable. We therefore analyze migration through four core activities that mirror the standard software life cycle (Figure~\ref{fig:se-activities}): \emph{Requirements Engineering}, \emph{Software Design}, \emph{Software Construction}, and \emph{Testing}. \emph{Requirements Engineering} dictates which original behaviors must be preserved and which platform-specific adaptations are permissible. \emph{Software Design} decomposes the migration task and defines the target structure, processing order, and component mappings. \emph{Software Construction} transforms source artifacts into target artifacts, iteratively repairing them based on environmental feedback. \emph{Testing} executes acceptance checks to verify whether the migrated result satisfies the migration contract. 

Crucially, these activities represent logical responsibilities rather than a rigid sequence: a system may interleave them, omit some entirely, or delegate them to human operators. Feedback loops frequently route back to earlier phases. Most acceptance failures stem from construction errors and thus trigger a return to the repair cycle within Software Construction. However, failures that localized code edits cannot resolve---such as flawed requirements, unsupported mappings, or unsuitable architectural decompositions---must instead be escalated back to Requirements Engineering or Software Design.

We assign each study according to its primary contribution. Studies focusing on a single activity---such as a novel requirements-recovery technique or a specific repair mechanism---are discussed within that respective activity. Conversely, studies proposing comprehensive systems that span multiple activities to deliver repository-level results are discussed under end-to-end migration. We further distinguish end-to-end \emph{development} (which reconstructs the source system into a new target implementation, typically across different languages or platforms) from end-to-end \emph{maintenance} (which modifies an existing codebase in place, usually to accommodate a new library, dependency, or language version). Within each individual activity, we categorize studies based on their focal artifact: Requirements Engineering by the format of the resulting specification; Software Design by the scope of architectural decisions; Software Construction by the transformed artifact and the extent of the repair loop; and Testing by the origin of the acceptance checks. Finally, the specific roles the model assumes within these activities are analyzed in Section~\ref{sec:role}, the orchestration of the workflow in Section~\ref{sec:agent}, and the provisioning of knowledge and feedback in Section~\ref{sec:technique}.

Platform migration exemplifies the concrete demands of each activity. In the context of mobile migration, for instance, navigation and lifecycle behaviors dictate the requirements; structural differences in system services and dependencies shape the design; permissions, threading, and data persistence must be meticulously handled during construction; and lifecycle- or device-dependent behaviors determine the specific conditions under which testing must execute. Consequently, we conclude each subsection by highlighting the remaining gaps that must be bridged to deliver a running application.

\subsection{Requirements Engineering}
\label{sec:se-spec}
\label{sec:spec}

Requirements Engineering determines which features and observable behaviors must be preserved, and which must be adapted for the target environment. Because migration tasks rarely begin with an explicit formal specification, the source implementation typically serves as the primary behavioral reference. However, it invariably embeds incidental design choices and legacy defects that should not be propagated. Consequently, we distinguish four forms of specification artifacts, ordered by how directly they can be verified against the migrated result.

\paragraph{Context descriptions}
The most prevalent form of specification collects descriptive information about the project. The
Android-to-iOS pilot study, for instance, builds a knowledge base from READMEs, comments,
issues, pull requests, and project websites, dynamically retrieving relevant entries during
translation~\cite{DBLP:journals/corr/abs-2507-16037}.
Such descriptions inform the model about the overarching purpose of the codebase, but they fail to
establish testable conditions for acceptance or failure. The same study
reports translated files that are syntactically correct yet entirely inappropriate for
the target platform even with this context available, demonstrating that understanding high-level intent
is insufficient for deriving an idiomatic target implementation.

\paragraph{Source design descriptions}
The second form recovers the structural architecture of the source system. RustPrint synthesizes
the source architecture into comprehensive documentation that records modules, data flows,
APIs, and design rationales, leveraging this documentation as a blueprint for
migration~\cite{DBLP:journals/corr/abs-2605-14634}.
A design description is more structured than general project context and can expose
architectural relationships that the target should preserve. RustPrint also compares source and target documentation and uses a RequirementRefiner to restore missing functionality. An explicit target behavior contract is still needed to specify acceptable platform adaptations.

\paragraph{Interface models}
The third form models the user interface. ArkTrans extracts an intermediate representation of
declarative UI files before mapping components to
ArkUI~\cite{zheng2026portingdeclarativeuiharmonyos},
and earlier work reconstructs a model of the GUI by observing its dynamic
execution~\cite{DBLP:conf/wcre/MemonBN03}.
An earlier iOS-to-Android porting effort similarly relied on
platform-independent descriptions of the application and its UI
structure~\cite{DBLP:journals/mis/LamhaddabLE19}.
Interface models enable direct comparisons against the target UI, yet they are strictly limited to
the structures or interactions explicitly observed. A static layout tree does not
capture every state transition, and an observed execution trace cannot reveal edge cases,
such as what occurs when a process is terminated and subsequently restored by the OS.

\paragraph{Supplied contracts}
The fourth form consists of specifications authored externally to the migration system,
typically as part of a benchmark. ScarfBench, for example, formally dictates the behaviors
that a migrated application must
exhibit~\cite{DBLP:journals/corr/abs-2605-06754}.
Such contracts often serve solely to evaluate the final result rather than to guide the
migration process: ECAT's benchmark provides feature checklists for each repository,
but explicitly withholds them from the migration agents, utilizing them exclusively for
evaluation~\cite{DBLP:journals/corr/abs-2608-09273}.
While supplied contracts can be verified directly, they shift the specification
burden from the automated system to the human task curators, and whether they actively
guide migration or merely evaluate it depends entirely on the experimental protocol.

Across all four forms, a recovered specification fundamentally describes what the source system
currently does, which does not always align with what the target system \emph{should} do. Observed source behavior inevitably includes legacy defects, workarounds specific to the source platform, or features that the target
platform handles via completely different paradigms. A rigorous migration specification must therefore
trace the provenance of each requirement, ensuring that an incidental source behavior is
not inadvertently elevated to a strict target requirement.

These four forms suggest a trade-off, although no reviewed study measures the effort involved: specifications that offer the most direct verifiability typically demand the highest manual preparation effort.
Some studies state individual allowances: ECAT does not require the exact Android UI structure to be reproduced~\cite{DBLP:journals/corr/abs-2608-09273}, and Mono2Sls replaces local authentication with a managed service~\cite{DBLP:journals/corr/abs-2604-24550}. However, none of the reviewed studies declares in advance a systematic, executable contract of which platform differences are deemed acceptable---such as determining whether a confirmed edit must survive a process restart while the accompanying screen transitions are allowed to differ. Migrating a complete application
critically depends on precisely this kind of nuanced specification, a gap that
Section~\ref{sec:challenge-platform} discusses in detail.
\subsection{Software Design}
\label{sec:se-design}
\label{sec:plan}

\emph{Software Design} dictates how the migration process is orchestrated, either prior to or during the generation of target artifacts. While isolated function-level translation requires minimal design, repository-level migrations demand explicit strategies for decomposing the workload, sequencing the processing steps, and establishing shared conventions across components. We categorize these design decisions by their scope, ranging from the localized ordering of translation units to the macro-architecture of the target system.

\paragraph{Decomposition and ordering}
Dependency analysis serves as the most prevalent foundation for decomposing a migration task. For instance, APIMig traces project dependencies to identify cascading chains of changes triggered by API evolution~\cite{DBLP:conf/ijcai/KuangXYYWKX25}. Similarly, PtrTrans and RustMap leverage dependency relations to partition projects into discrete translation units and dictate their processing sequence~\cite{DBLP:journals/pacmse/YuanMCSWLP26, DBLP:conf/iceccs/CaiLHYWLWYJ25}, while DepWareTrans groups interdependent files into batches, enabling incremental repository migration across co-executable languages~\cite{DBLP:journals/corr/abs-2608-14128}. Collectively, these studies utilize dependency graphs for three interconnected purposes: identifying code blocks that must co-evolve, defining the atomic units of translation, and determining the optimal processing order and batch boundaries.

\paragraph{Interfaces and skeletons}
A second category of design decisions establishes shared conventions prior to implementation, ensuring that independently translated units integrate seamlessly. For example, TyMCRaT orders functions via the call graph, generating and validating candidate Rust signatures before translating the function bodies~\cite{DBLP:journals/ese/HongR25}. Oxidizer decouples the verification of type compatibility from semantic validation~\cite{DBLP:journals/pacmpl/ZhangDWPK25}; EvoC2Rust preemptively constructs a compilable project skeleton~\cite{DBLP:conf/icse-seip/WangYSWCZSXG26}; and His2Trans reconstructs the target structure using build information~\cite{wang2026his2transknowledgeguidedagenticframework}. By resolving types, identifiers, and interfaces upfront, these approaches mitigate integration errors that would otherwise remain latent until the units are assembled. It is important to note, however, that while a compilable skeleton guarantees structural consistency, it does not inherently guarantee behavioral preservation.

\paragraph{Target architecture and translation paths}
A third group of decisions governs the macro-level organization of the target system or the trajectory of the transformation process itself. RustPrint utilizes its synthesized architectural documentation to structure target modules~\cite{DBLP:journals/corr/abs-2605-14634}, whereas Mono2Sls defines function boundaries, communication protocols, and managed-service substitutions when migrating from a monolithic to a serverless architecture~\cite{DBLP:journals/corr/abs-2604-24550}. Crucially, these decisions often reshape module boundaries rather than strictly mirroring the source structure. From a routing perspective, InterTrans optimizes the transformation path by searching for optimal sequences of intermediate languages to bridge the source and target~\cite{DBLP:conf/icse/Macedo00CA25}.

\paragraph{Mappings}
The final category of design decisions involves mapping source constructs to their appropriate target counterparts. In UI migration, ArkTrans employs heuristics to explicitly map source UI components to ArkUI elements prior to code generation~\cite{zheng2026portingdeclarativeuiharmonyos}. Conversely, the majority of other studies rely on implicit mappings latent within the model's output. Furthermore, very few studies systematically address how to handle scenarios where the target platform lacks a direct equivalent, though the Android-to-iOS pilot study notably highlights package-level translation failures stemming from incompatible third-party libraries, divergent platform-specific designs, and distinct data storage mechanisms~\cite{DBLP:journals/corr/abs-2507-16037}.

Naturally, the complexity of design requirements scales with the granularity of the migration task. At the module level, design concerns are largely confined to the selected unit and its immediate interfaces. At the repository level, the design must carefully coordinate distinct translation units, shared types, cross-module dependencies, and build system integration. Finally, where the repository must run as an application, the design must additionally orchestrate the holistic assembly of components and the substitution of platform-specific services, akin to the managed-service replacements in Mono2Sls. Currently, design techniques are overwhelmingly concentrated at the repository level; architectural decisions that arise only for a running application---such as how to systematically replace a platform service lacking a direct target counterpart---remain a critical, yet largely unaddressed, frontier.

\subsection{Software Construction}
\label{sec:se-construction}
\label{sec:execute}

\emph{Software Construction} encompasses the generation of target artifacts and their subsequent refinement. As the most extensively studied activity in the migration literature, it can be broadly divided into transformation (the initial production of artifacts) and repair (their iterative revision via feedback).

\subsubsection{Transformation}
\label{sec:se-transformation}

We classify transformation techniques based on the specific target artifact they modify. The exact mechanisms through which the model participates in this process---such as generating code directly versus synthesizing rules for a deterministic engine to execute---are analyzed in Section~\ref{sec:role}.

\paragraph{Program logic}
The vast majority of studies focus on transforming program logic, encompassing language constructs, type systems, and memory handling. Early neural approaches primarily translated individual isolated functions~\cite{DBLP:conf/nips/RoziereLCL20}. Guess \& Sketch combines language-model assembly translation with alignment and confidence estimates and symbolic-solver repair~\cite{DBLP:conf/iclr/LeeMKC0CWR24}. More recent approaches train models using execution-based rewards for function-level translation~\cite{DBLP:journals/corr/abs-2607-19966}. When the source and target languages possess fundamentally divergent type and memory models, logic transformation must also explicitly adapt types and pointers. In C-to-Rust migration, for instance, various approaches utilize data-flow information to infer appropriate Rust types for global variables and pointers~\cite{DBLP:conf/soap-ws/XuH25}; rewrite raw pointers into safer, idiomatic forms~\cite{DBLP:journals/corr/abs-2505-04852}; and systematically reduce the unsafe code blocks generated by legacy rule-based transpilers~\cite{DBLP:journals/tse/NitinKVR26}. Mobile migration further complicates this by altering the language and the runtime simultaneously: in Android-to-iOS migration, methods are first translated to Swift using syntax trees and retrieved project context, and only subsequently assembled into cohesive classes and components~\cite{DBLP:journals/corr/abs-2507-16037}.

\paragraph{APIs and libraries}
A second category focuses on replacing API and library usages with their target-platform or upgraded counterparts. Direct substitution methods, such as GUPPY, automate the update of deprecated Android API usages, evaluating performance across 360 instances of 20 deprecated APIs and 156 usages from API levels 33 and 34~\cite{DBLP:journals/tse/MahmudDCYNY26}. Conversely, a growing body of work focuses on synthesizing reusable transformations for library migration. MELT leverages an LLM to generate and validate migration examples---supplementing those mined from pull requests---from which a secondary algorithm infers generalized rewrite rules~\cite{DBLP:conf/kbse/RamosMLMMG23}; similarly, Allain et al.\ evaluate rules synthesized directly from before-and-after examples for API and language-version upgrades~\cite{DBLP:journals/corr/abs-2609-03592}. Extending this paradigm, SPELL composes programmatic migration scripts for Python libraries~\cite{DBLP:journals/corr/abs-2602-01107}, while BigBag generates robust AST transformations that adapt Java clients to breaking dependency updates, subsequently transferring these rules to other affected client projects~\cite{DBLP:journals/corr/abs-2606-24446}. While reusable transformations apply wherever their structural pattern matches, they are inherently limited by the completeness of the usage patterns they cover: SPELL reports incomplete migrations when scripts fail to account for obscure API usages, and BigBag notes that cross-project transfer heavily depends on how uniformly different clients invoke the affected API~\cite{DBLP:journals/corr/abs-2602-01107, DBLP:journals/corr/abs-2606-24446}.

\paragraph{User interfaces}
A third group transforms user interfaces, a task that in mobile migration frequently necessitates changing both the underlying framework and the programming language. GUIMigrator transpiles Android XML layouts to Compose and SwiftUI utilizing deterministic, semantics-preserving rules~\cite{DBLP:journals/pacmse/GaoHYX26}. At the LLM level, UITrans translates Android UIs to HarmonyOS ArkUI at the component, page, and project level, parsing pages and the interactions between main and sub-pages and assembling the result~\cite{DBLP:conf/internetware/GongWCHW25}, whereas ArkTrans populates a partial ArkUI skeleton from Compose or SwiftUI sources using a one-shot demonstration, followed by pattern-based structural refinement~\cite{zheng2026portingdeclarativeuiharmonyos}. ArkTrans operates on independent pages and explicitly excludes external-file and third-party dependencies from its evaluation, whereas UITrans handles dependencies between pages within the UI subsystem. Neither study shows that the behavior behind the UI, such as state management and data flows, is preserved.

\paragraph{Build, configuration, and resources}
A fourth group addresses the foundational artifacts required to make the codebase buildable and executable: build scripts, dependency manifests, multimedia resources, and platform-specific configurations. His2Trans extracts build information to reliably reconstruct the target project structure~\cite{wang2026his2transknowledgeguidedagenticframework}; MigrationBench's maximal difficulty setting mandates that project dependencies be upgraded in tandem with the core language version~\cite{DBLP:conf/kdd/LiuLZCLNOSHTD26}; and ECAT bypasses structural generation altogether by initiating migration from a predefined target template~\cite{DBLP:journals/corr/abs-2608-09273}. Ultimately, very few studies currently treat these configuration artifacts as primary objects of automated transformation; instead, they are typically bypassed via predefined templates or extensive manual preparation.

\paragraph{Assembly}
Finally, transformed artifacts must be meticulously assembled into a coherent, functioning codebase. AlphaTrans translates program fragments in isolation before systematically recomposing them into a unified target repository~\cite{DBLP:journals/pacmse/IbrahimzadaKPAPSJ25}, while EvoC2Rust incrementally replaces the function stubs within its pre-generated skeleton~\cite{DBLP:conf/icse-seip/WangYSWCZSXG26}. As established previously, \emph{Software Design} dictates the structural skeleton and processing order (Section~\ref{sec:se-design}); assembly executes the physical integration and resolves any residual inconsistencies that surface during combination.

In summary, current transformation research remains disproportionately concentrated on program logic and APIs, which can be conveniently evaluated at the module level. While user interface migration has garnered growing attention, the holistic transformation of build files, configurations, and resources---which are indispensable for delivering a truly working application---is rarely undertaken by the automated migration systems themselves.

\subsubsection{Repair}
\label{sec:se-repair}
\label{sec:verify-fix}

Repair revises a candidate after a check, during construction or testing,
reveals an error. We distinguish repair by how far a fix must reach: within a
single unit or across several units. We then discuss the failures that repair
cannot resolve and that must be escalated to an earlier activity. The signals
that drive repair, such as compiler diagnostics, test failures, and model
judgments, are discussed in Section~\ref{sec:technique}.

\paragraph{Unit-local repair}
The simplest repair corrects an error within a single unit, such as a
compilation or type error in one function. Byam repairs Java projects whose
compilation breaks after a dependency update, using compiler diagnostics and
API differences as context and rerunning the build
afterwards~\cite{DBLP:journals/ese/ReyesMBNBM26},
and LLMigrate applies compiler-guided repair to C functions translated for
selected Linux kernel
modules~\cite{DBLP:journals/corr/abs-2503-23791}.
GUPPY refines API updates when generated tests
fail~\cite{DBLP:journals/tse/MahmudDCYNY26}.

\paragraph{Cross-unit repair}
A harder kind of repair addresses errors that arise between units: mismatched interfaces,
references to definitions that were renamed or not yet translated, and
callers that no longer agree with their callees. Fixing them requires
changing several units consistently. TyMCRaT supplies translated callee
interfaces to callers and repairs the resulting type
errors~\cite{DBLP:journals/ese/HongR25},
and the Android-to-iOS pilot study checks references across files against the
source dependency graph, feeding the detected problems back into further
translation
rounds~\cite{DBLP:journals/corr/abs-2507-16037}.
Internal reference mismatches are nevertheless the largest failure category
in that study, which indicates how difficult cross-unit consistency is to
restore after the fact.

\paragraph{Escalation to earlier activities}
Some failures originate upstream of construction. If the target platform
lacks a capability, a mapping is wrong, or the decomposition separates code
that must change together, repeated code edits may suppress the symptom
without addressing the cause. Such failures should be escalated to
requirements engineering or software design (Figure~\ref{fig:se-activities}).
Among the systems we reviewed, we found none
that detects this situation and escalates it.

Repair loops also differ in when they stop. Most stop after a fixed number of
rounds or once the check passes; the Android-to-iOS pilot study, for example,
runs three refinement rounds per file and reports diminishing returns from
further
iterations~\cite{DBLP:journals/corr/abs-2507-16037}.
ECAT instead accepts a change only if it strictly lowers its code entropy and
stops once the entropy stays below a small threshold for consecutive
iterations, which ties both acceptance and stopping to a measured
quantity~\cite{DBLP:journals/corr/abs-2608-09273}.
A fixed budget bounds cost but may stop before an error is fixed, whereas an
acceptance rule tied to a measured quantity can stall when the measure no
longer reflects the remaining errors.

Repair loops are therefore effective for unit-local errors and partly
effective for cross-unit errors, but they do not recognize failures that need
escalation. General studies of
self-correction show that a model can revise its output using its own
feedback or execution
results~\cite{DBLP:conf/iclr/ChenLSZ24, DBLP:conf/nips/MadaanTGHGW0DPY23},
but whether additional rounds help depends on what the feedback can reveal.
In a running application, many errors concern interactions between modules
and platform behavior, which compiler and unit-level checks rarely expose.

\subsection{Testing}
\label{sec:se-testing}
\label{sec:test}

Testing decides whether the migrated result satisfies the migration contract.
Its central problem in migration is obtaining acceptance checks for the
target, since tests written for the source often depend on source APIs or
platforms. We classify testing by the origin of the acceptance checks: tests
carried over from the source, newly generated tests, or the source program
itself used as a reference. When a failed check is used to revise the
candidate, the revision belongs to repair (Section~\ref{sec:se-repair}); the check itself still counts as testing here, whether it feeds repair or serves as final acceptance.

\paragraph{Carried-over source tests}
When the source project has tests, they can be carried over to the target.
At the code level, tests are translated along with the code. AlphaTrans
executes source tests in isolation and also translates them to validate the
recomposed target
repository~\cite{DBLP:journals/pacmse/IbrahimzadaKPAPSJ25},
and RustPrint translates source tests for its C-to-Rust
migration~\cite{DBLP:journals/corr/abs-2605-14634}.
Translated tests inherit the coverage of the source suite, and they may be
translated by the same model as the code, so a shared misunderstanding can
affect both.

At the GUI level, tests are migrated between applications or platforms, which
comes closest to accepting a running application. TestMig migrates GUI test
cases from iOS to
Android~\cite{DBLP:conf/issta/QinZW19}. A more
recent approach executes the source test to enrich its events with widget
information, abstracts the test into natural language, and transfers it
through interaction with the target
application~\cite{DBLP:journals/pacmse/BeyzaeiTRMM25}.
Its assertions can check visible widgets, expected text, and UI state. Correctness of hidden state or persisted data requires suitable additional oracles. For Android-to-OpenHarmony migration, an
empirical study adapts existing test migration methods to the target device
layer and finds that most failed migrations are linked to characteristics
specific to
OpenHarmony~\cite{DBLP:journals/corr/abs-2607-11245}.
The authors then redesign how test intentions are extracted, how GUI content
is read, and how actions are executed on the target, showing that adapting
only the device layer is not enough when the UI framework of the platform
differs.

\paragraph{Generated tests}
Where source tests are insufficient, new tests can be generated. ReCodeAgent
generates tests guided by coverage in addition to translating existing
ones~\cite{DBLP:journals/corr/abs-2604-07341},
and GUPPY generates tests to check API
updates~\cite{DBLP:journals/tse/MahmudDCYNY26}.
Generated tests extend coverage beyond the source suite, but their expected
results must themselves be correct, which is difficult to establish without a
specification.

\paragraph{Differential checks}
A third approach uses the source program itself as the reference. C2RustXW
combines compilation with differential execution of the source and target
programs~\cite{DBLP:journals/corr/abs-2603-28686}.
Differential checks need no hand-written expected outputs, but they cover
only the inputs that are executed and, when outputs are compared exactly, assume that the source behavior should
be preserved exactly, which conflicts with platform differences that are
intended.

The three origins thus trade the availability of checks against their
independence: carried-over tests exist only where the source has them,
generated tests depend on the correctness of their own expectations, and
strict differential checks inherit every behavior of the source. GUI-level test
migration brings acceptance closest to user-visible behavior, but in current
work it is a separate task from implementation migration: it assumes that the
target application already exists. Combining the two, so that migrated tests
accept a migrated application, is one promising route to accepting a migrated
application, alongside acceptance by people on the target platform and newly generated tests whose expectations are verified.

\subsection{End-to-End Software Development}
\label{sec:se-dev}
\label{sec:end-to-end}

End-to-end development systems rebuild a source system as a new target
implementation by combining several activities. We describe each system by
the activities it covers, how they are connected, and where the workflow
stops, and organize the systems by the granularity of their result.

\paragraph{Repository-level systems}
AlphaTrans illustrates a repository-level workflow that covers software
design, software construction, and testing. It builds a compilable target skeleton, translates
program fragments in reverse call order with relevant context, recomposes the
project, and validates it with both source and translated
tests~\cite{DBLP:journals/pacmse/IbrahimzadaKPAPSJ25}.
ReCodeAgent distributes similar responsibilities across an analyzer, a
planner, a code and test translator, and a validator, and adds
coverage-guided test generation, so that testing is part of the workflow
rather than a final
step~\cite{DBLP:journals/corr/abs-2604-07341}.
RepoTransAgent is repository-aware in a narrower sense: it translates one function at a time using context from existing projects, so its output unit is the
function~\cite{DBLP:journals/corr/abs-2508-17720}.
A cluster of C-to-Rust systems follows the same pattern of dependency-aware
design, unit translation, and project-level assembly. RustMap and EvoC2Rust
emphasize dependency-guided composition and skeleton-based integration,
respectively; RustPrint adds a documented design blueprint, documentation-guided functional refinement, and translated
tests; and His2Trans adds build-derived structure and historical migration
knowledge~\cite{DBLP:conf/iceccs/CaiLHYWLWYJ25, DBLP:conf/icse-seip/WangYSWCZSXG26, DBLP:journals/corr/abs-2605-14634, wang2026his2transknowledgeguidedagenticframework}.
Across these systems, requirements engineering is the activity most often
absent: the
source code and its tests serve as the de facto specification.

\paragraph{Systems that must deliver a running application}
Three systems carry the deployed-application requirement (Table~\ref{tab:deployed}). The Android-to-iOS pilot study combines
requirements engineering, software design, and software construction in one
workflow~\cite{DBLP:journals/corr/abs-2507-16037}.
A specification extraction agent builds the project knowledge base, a code
translation agent translates methods bottom-up along the dependency graph and
assembles classes, components, and the project, and a code validation agent
checks references, syntax, and style and feeds problems back for further
rounds. This validation raises the share of valid files from 43.2\% to
70.7\% and removes all detected syntax errors. None of the workflow's checks executes code: the translated projects could
not be built without substantial manual effort, so the study evaluates the
validity of individual files, judged by the authors, rather than the
behavior of the application. Mono2Sls migrates a monolith to a serverless
architecture through a multi-stage pipeline with static analysis, in which
design decisions about function boundaries and managed services are central
to the
result~\cite{DBLP:journals/corr/abs-2604-24550}. Inside its workflow, a validator checks the generated artifacts and deployment templates statically; cloud deployment and behavioral tests appear only in the paper's evaluation.

ECAT migrates Android repositories to HarmonyOS through an iterative
loop~\cite{DBLP:journals/corr/abs-2608-09273}.
Starting from the source repository and an initial target template, a
discriminator with an independent context estimates a \emph{code entropy}
and returns a structured ``text gradient'' that lists defects and prioritized
repair targets, and a generator edits the repository accordingly. The code
entropy aggregates fourteen dimensions: twelve static ones, such as
compilation, parity of permissions and the data layer with the source, and
navigation entry points, and two dynamic ones, namely screenshot alignment
judged by a multimodal model and cold-start and smoke-click probes on an
emulator. A change is accepted only if it lowers the entropy, and knowledge
distilled from entropy-reducing trajectories is stored in a shared memory
tree for later reuse. ECAT has no separate planning stage; decomposition
happens within each iteration through the prioritized work items of the text
gradient. In terms of our framework, ECAT concentrates on software
construction:
transformation and repair are tightly coupled, no specification is recovered,
and the runtime probes serve as repair signals rather than as a separate
acceptance activity. The feature checklists used to evaluate ECAT are
withheld from the migration agents, which keeps evaluation independent of
the repair loop.

These three systems show a common pattern. As the delivered
unit grows, systems cover more of construction and repair but less of testing:
none of them includes a final acceptance step, held out from the repair loop, that executes the migrated
application against its required behavior.

\subsection{End-to-End Software Maintenance}
\label{sec:se-maint}

End-to-end maintenance systems migrate an existing codebase in place, keeping
its language and repository while adapting it to a new library, dependency,
or language version. We organize them by what triggers the migration, from
local API changes to organization-wide programs.

\paragraph{Library and API replacement}
Library migration replaces one library or library version with another
throughout a codebase. Studies of migrating Python applications to a new
version of SQLAlchemy evaluate LLMs and, more recently, an agent mode of a
coding assistant on this
task~\cite{DBLP:conf/esem/AlmeidaXV24, DBLP:conf/agent/AlmeidaVX26},
and a broader study reports that GPT-4o correctly performs 94\% of the evaluable code
changes required by Python library
migrations (1,867 of 1,996, with a further 914 changes excluded); at the level of whole migrations, 179 of 314 (57\%) are fully correct, and 64\% (16 of 25) pass the tests on the subset whose tests the developers' own migration passes~\cite{DBLP:journals/corr/abs-2504-13272}. The three figures have different denominators.

\paragraph{Dependency updates}
A dependency update can break code that compiled before. Byam addresses this
case for Java projects: it collects compilation errors and API differences
after an update, asks the model for a fix, and rebuilds and reruns the tests
until the project compiles
again~\cite{DBLP:journals/ese/ReyesMBNBM26}.

\paragraph{Language and framework versions}
Upgrading a language or framework version usually affects the whole
repository. Benchmarks for Java version upgrades and for cross-framework
migration of enterprise Java applications evaluate general-purpose coding
agents on such
tasks~\cite{DBLP:conf/kdd/LiuLZCLNOSHTD26, DBLP:journals/corr/abs-2510-04852, DBLP:journals/corr/abs-2605-06754}.
The agents receive the repository and a target version or framework and must
deliver a project that builds and passes its tests; we discuss the benchmarks
themselves in Section~\ref{sec:eval-resources}.

\paragraph{Large-scale migration programs}
Organizations also run migrations across many repositories. At Google, a
workflow combines automated discovery of the locations that need to change
with LLM-generated edits; over a year-long program, the LLM generated 74.45\%
of the submitted
changes~\cite{DBLP:conf/sigsoft/ZiftciNSKCK25}. Developers
estimated that the total migration time fell by half compared with earlier
manual migrations. CodePlan frames
repository editing as planning: it combines incremental dependency analysis
and change may-impact analysis to derive a chain of edits, and evaluates the
approach on package migration in C\# and temporal edits in
Python~\cite{DBLP:journals/pacmse/BairiSKCIPRAS24}.
In this evaluation, five of six repositories (three for C\# package migration and three for Python temporal edits) pass the validity checks, which are build and static type checks,
whereas baselines without planning pass none.

In-place migration imposes requirements that rebuilding does not. First,
every change can affect code that was not edited, so the system must find
all affected locations, which is the role of change-location discovery at
Google and of may-impact analysis in CodePlan. Second, the existing tests
encode the expected behavior and must not be weakened to make the migration
pass, which motivates benchmark safeguards such as preserving test identifiers, preventing reductions in executed test counts, and limiting coverage loss~\cite{DBLP:conf/kdd/LiuLZCLNOSHTD26,DBLP:journals/corr/abs-2510-04852}. Third, changes in a shared codebase are reviewed
and merged incrementally, so a migration must be divisible into changes that
are individually correct. Maintenance-oriented migration benefits from an
existing build and test suite, which should make its results easier to check
automatically than those of development-oriented migration; in return, its
tasks are narrower and rarely involve a change of platform.

\begin{table*}[t]
\caption{Activities covered by representative studies. Category: R, D, C, T =
single-activity study of requirements engineering, software design, software
construction, or testing;
Dev = end-to-end development; Maint = end-to-end maintenance. NR: not
reported in the cited account. The last column lists checks run inside the workflow, whether they feed repair or serve as final acceptance. The table is a selection, not a coding of all Class~A studies, and includes one Class~B study on test migration.}
\label{tab:stage-artifacts}
\centering\scriptsize
\setlength{\tabcolsep}{3pt}
\begin{tabular}{p{0.11\textwidth}p{0.05\textwidth}p{0.16\textwidth}p{0.18\textwidth}p{0.22\textwidth}p{0.18\textwidth}}
\toprule
Study & Cat. & Requirements engineering & Software design & Software construction & Testing \\
\midrule
ArkTrans~\cite{zheng2026portingdeclarativeuiharmonyos} & C & Extracted UI representation & Component mapping; partial skeleton & Skeleton completion; pattern-based post-fixing & NR \\
TyMCRaT~\cite{DBLP:journals/ese/HongR25} & D & NR & Call-graph order; candidate signatures & Body translation; type repair from compiler feedback & Compiler feedback \\
GUPPY~\cite{DBLP:journals/tse/MahmudDCYNY26} & C & NR & NR & Updated API usages; test-driven refinement & Generated tests \\
Android-to-OpenHarmony test study~\cite{DBLP:journals/corr/abs-2607-11245} & T & Source test intent & NR & NR & Migrated GUI tests on existing apps \\
AlphaTrans~\cite{DBLP:journals/pacmse/IbrahimzadaKPAPSJ25} & Dev & Source code and tests & Skeleton; reverse call order & Fragment translation; recomposition & Source and translated tests \\
ReCodeAgent~\cite{DBLP:journals/corr/abs-2604-07341} & Dev & NR & Planner & Code translation; validator feedback & Translated and coverage-guided tests \\
C2RustXW~\cite{DBLP:journals/corr/abs-2603-28686} & Dev & NR & Dependency-aware unit order & Multi-file translation guided by program structure & Compilation; differential execution \\
RustPrint~\cite{DBLP:journals/corr/abs-2605-14634} & Dev & Architecture documentation; source--target documentation comparison & Module plan & Modular implementation; functional refinement & Translated source tests \\
Pilot study~\cite{DBLP:journals/corr/abs-2507-16037} & Dev & Project knowledge base & Bottom-up dependency order & Method-to-project translation; reference, syntax, and style repair & Reference, syntax, and style validation; nothing executed \\
Mono2Sls~\cite{DBLP:journals/corr/abs-2604-24550} & Dev & NR & Function boundaries; managed services & Multi-stage transformation with static analysis & Static consistency and deployment-template validation \\
ECAT~\cite{DBLP:journals/corr/abs-2608-09273} & Dev & NR (evaluation checklists withheld) & Per-iteration prioritized work items & Generator--discriminator loop; memory tree & Compilation, static parity, screenshot judgment, emulator probes; all feed repair \\
Byam~\cite{DBLP:journals/ese/ReyesMBNBM26} & Maint & NR & NR & Patches from compiler errors and API differences & Build and test rerun \\
CodePlan~\cite{DBLP:journals/pacmse/BairiSKCIPRAS24} & Maint & Task instructions & Dependency and may-impact analysis; adaptive plan & Chain of LLM edits & Build and static type checks \\
Google~\cite{DBLP:conf/sigsoft/ZiftciNSKCK25} & Maint & NR & Change-location discovery & LLM-generated edits & Build and tests, then human review \\
\bottomrule
\end{tabular}
\end{table*}

Table~\ref{tab:stage-artifacts} summarizes the activities covered by
representative studies; the three observations that answer RQ2 are qualitative and refer to these systems. First,
requirements engineering is the weakest activity among them: most systems treat the
source code and
its tests as the specification, and the explicit specifications that exist
are either descriptive or supplied by people. Second, software construction is the
most developed activity, but its transformations concentrate on program logic
and APIs, and its repair loops handle local errors far better than errors
between units or errors that originate upstream. Third, testing rarely
reaches a running application, and end-to-end systems cover less of testing
as the delivered unit grows. How models take part in these activities and how
their work is coordinated are the subjects of the next two sections.

%% file: sections/role_perspective.tex
\section{Role-Based Perspectives}
\label{sec:role}

Section~\ref{sec:se} examines the software engineering activities supported by LLM-assisted migration systems, but the same activity may be performed by different actors. For example, software construction may involve an LLM directly generating target code, synthesizing transformation rules for a deterministic engine, or directing tools to inspect and validate the resulting implementation. These approaches differ in the responsibilities assigned to the model, even when they contribute to the same migration activity.

To address the role dimension of RQ3, we distinguish two complementary roles: \textit{LLM-as-Editor} and \textit{LLM-as-Orchestrator}. An LLM acts as an Editor when it directly authors or modifies migration artifacts, including target code, corrective patches, and reusable transformation rules. It acts as an Orchestrator when it directs migration activities beyond artifact production, including context acquisition, task planning, tool invocation, and feedback interpretation. These roles are not mutually exclusive: a model may both produce migration artifacts and coordinate the operations surrounding their development and validation.

We classify roles according to the responsibilities demonstrably assigned to the LLM, rather than the capabilities available to the overall system. In particular, the presence of a compiler or test runner does not establish that the model orchestrates its execution. A deterministic workflow may invoke these tools at predetermined stages and merely forward their outputs to the model. We therefore distinguish model-directed operations from those implemented by deterministic components or human operators.

\subsection{LLM-as-Editor}
\label{sec:se-editor}

An LLM acts as an Editor when it directly produces or modifies artifacts that implement the migration. This role covers not only the generation of target-language code but also the refinement of existing candidates and the synthesis of reusable transformation logic. We distinguish three forms according to the artifacts produced.

\subsubsection{Direct Code Generation and Translation}

The most straightforward form uses an LLM to transform source artifacts into target-language or target-platform implementations. The model directly authors the resulting code, which may subsequently undergo structural refinement, integration, and validation. In mobile UI migration, UITrans addresses the translation of Android UI implementations into HarmonyOS interfaces~\cite{DBLP:conf/internetware/GongWCHW25}, while heuristic-guided approaches investigate LLM-assisted porting of declarative UI to HarmonyOS~\cite{zheng2026portingdeclarativeuiharmonyos}. GUIMigrator, by contrast, performs rule-based, semantics-preserving transpilation of Android XML interfaces into Compose and SwiftUI without a model in the Editor role~\cite{DBLP:journals/pacmse/GaoHYX26}. At the repository level, AlphaTrans uses LLM-based translation within a neuro-symbolic compositional framework to produce target-language code fragments that are subsequently recomposed into a repository-level result~\cite{DBLP:journals/pacmse/IbrahimzadaKPAPSJ25}. Similarly, RustMap combines program analysis with LLM-based C-to-Rust translation at project scale~\cite{DBLP:conf/iceccs/CaiLHYWLWYJ25}. Although these approaches differ in their target artifacts and integration mechanisms, the Editor role concerns the model's contribution to producing the migrated implementation.

\subsubsection{Code Repair and Refinement}

Editing also occurs after an initial transformation, when compilation errors, dependency incompatibilities, or test failures reveal defects in the generated implementation. In this setting, the LLM receives diagnostic information and produces revised code or corrective patches. Byam addresses breaking dependency updates by combining compiler diagnostics and API differences with LLM-generated patches, followed by subsequent build or test operations~\cite{DBLP:journals/ese/ReyesMBNBM26}. Automated updates of deprecated Android API usages similarly investigate the use of LLMs to generate replacements for outdated API calls~\cite{DBLP:journals/tse/MahmudDCYNY26}. These approaches differ from one-pass translation because the model modifies an existing candidate in response to identified problems. Importantly, generating a corrective patch is an Editor responsibility, whereas invoking the compiler or test suite is a separate process responsibility considered under the Orchestrator role.

\subsubsection{Transformation Rule and Script Synthesis}

A further form of editing produces reusable transformation logic rather than the migrated code itself. Instead of independently translating every affected code fragment, an LLM can synthesize rules or scripts that a deterministic engine subsequently applies. Allain et al. investigate the synthesis of code transformation rules from examples, examining both the potential and limitations of LLM-generated transformation logic~\cite{DBLP:journals/corr/abs-2609-03592}. This separates the authorship of migration logic from its execution: the LLM produces the transformation mechanism, while a conventional engine applies the corresponding changes to source code. We include this form of work because the model directly authors an artifact that implements the migration, even though the artifact is not itself the final target program.

These three forms differ in the nature and reusability of their outputs. Direct translation produces target implementations, repair modifies existing candidates, and rule synthesis produces reusable transformation mechanisms. They should therefore be distinguished when analyzing the contributions of LLMs to software construction. In particular, generating a reusable transformation rule should not be treated as equivalent to successfully migrating the projects to which that rule is applied.

\subsection{LLM-as-Orchestrator}
\label{sec:sub-llm-orchestrator}

An LLM acts as an Orchestrator when it directs migration activities beyond directly producing or modifying artifacts. Rather than responding only to a predefined code transformation request, the model participates in determining what information is needed, which operations should be performed, and how their results should influence subsequent migration steps. This role is particularly relevant to agentic migration systems, where the model interacts with repositories, external tools, and execution environments.

\subsubsection{Context Acquisition and Analysis}

Migration requires information distributed across source files, documentation, dependency declarations, and project configurations. An LLM may direct the acquisition and examination of such information to establish the context needed for subsequent migration decisions. The Android-to-iOS pilot study investigates an agentic translation workflow that incorporates specification extraction and project-level context into the migration process~\cite{DBLP:journals/corr/abs-2507-16037}. Documentation-guided agentic migration from C to Rust similarly explores the use of project documentation to support codebase-level translation~\cite{DBLP:journals/corr/abs-2605-14634}. These approaches illustrate how the model's responsibilities may extend beyond translating an isolated code unit to examining information relevant to the broader project. However, when context is collected exclusively through predetermined retrieval or static-analysis procedures, the corresponding operations should not automatically be attributed to the LLM.

\subsubsection{Planning and Task Decomposition}

At larger migration granularities, the model may also participate in determining how migration work should be divided and executed. This includes identifying migration units, establishing dependencies, prioritizing tasks, and determining subsequent operations. ReCodeAgent organizes repository-level translation and validation through specialized agents with distinct responsibilities~\cite{DBLP:journals/corr/abs-2604-07341}. The Android-to-iOS pilot study also illustrates the use of separate migration stages for specification extraction, translation, and validation~\cite{DBLP:journals/corr/abs-2507-16037}. These approaches demonstrate how migration responsibilities can extend from local code generation to coordinating multiple activities across a project. Nevertheless, the presence of predefined stages does not by itself establish model-directed planning. Whether the next task is selected by the LLM or prescribed by program code must be examined separately.

\subsubsection{Tool Invocation and Execution}

A further responsibility involves directing external tools to perform migration-related operations. These include repository exploration, dependency inspection, static analysis, compilation, test execution, and interaction with target environments. The model may select an appropriate tool, formulate its inputs, invoke it, and use the returned information to guide subsequent operations. Agentic repository migration systems such as ReCodeAgent provide examples of workflows in which LLM-based agents participate in translation and validation beyond isolated code generation~\cite{DBLP:journals/corr/abs-2604-07341}. Documentation-guided C-to-Rust migration likewise investigates an agentic workflow that operates at codebase level~\cite{DBLP:journals/corr/abs-2605-14634}. However, the existence of tools in a migration pipeline does not necessarily mean that the LLM controls their execution. A system in which program code invokes a compiler at predetermined stages and forwards the diagnostics to the model represents a different division of responsibilities from one in which the model selects and initiates the compilation operation.

\subsubsection{Feedback Interpretation and Iterative Control}

Orchestrating migration also involves interpreting the results of previous operations and determining whether the current candidate should be revised, further investigated, or advanced to another stage. For example, agentic migration workflows may incorporate validation results into subsequent translation or repair decisions~\cite{DBLP:journals/corr/abs-2604-07341}. In such cases, the model's contribution extends beyond generating an artifact to deciding how the migration process should respond to observed problems. This responsibility should be distinguished from merely receiving diagnostic messages and generating a patch: the latter demonstrates the Editor role, whereas interpreting feedback and directing subsequent operations demonstrates the Orchestrator role. Furthermore, model-generated judgments should not be equated with independent proof of correctness. The evidence used to accept a migration result is examined separately in Section~\ref{sec:eval}.

The Orchestrator role therefore concerns the model's responsibility for directing the activities surrounding migration artifact production. Its scope ranges from acquiring project context and decomposing migration tasks to invoking tools and responding to execution feedback. The degree of responsibility varies across systems: some models participate only in selected planning or analysis steps, while others direct successive operations throughout the migration process.

Taken together, the two roles describe complementary dimensions of LLM participation in migration. The Editor role concerns the authorship and modification of migration artifacts, whereas the Orchestrator role concerns directing the activities through which those artifacts are developed and assessed. A system may exhibit either role or both. Of the 20 systems whose roles we coded (Table~\ref{tab:agent-architectures}), nine confine the model to the Editor role and eleven also give it Orchestrator responsibilities, four of the latter coded from the authors' own description. The architectural question of how these responsibilities are organized and controlled---through predefined workflows, a single agent, or multiple agents---is examined separately in Section~\ref{sec:agent}.

%% file: sections/agent_perspective.tex
\section{Agent Perspectives}
\label{sec:agent}

While Section~\ref{sec:role} examines the roles assumed by LLMs, this section investigates how migration activities are organized and controlled. In particular, we examine who determines the next action during migration. An LLM-based system may involve multiple model calls, external tools, and iterative feedback without necessarily exhibiting agentic control. A workflow may incorporate extensive program analysis and repeated model interactions while retaining a fixed execution order.

Following the agent definition in Section~\ref{sec:bg-llm}, we distinguish three orchestration structures according to their control locus. An \textit{agentless workflow} is controlled by program code, which determines the execution steps, their order, retry conditions, and stopping rules, while the LLM generates content at designated stages. A \textit{single-agent system} allows one LLM instance to select actions from an available tool space and decide when to stop. A \textit{multi-agent system} employs two or more LLM instances with distinct roles and separate contexts that exchange intermediate artifacts, each selecting its own actions; an LLM that is merely called at a fixed point to produce content is a helper, not an agent. The hand-offs between these agents may themselves be fixed by program code; we record such scheduling as a subtype of multi-agent orchestration rather than a separate category.

We apply two coding rules throughout this section. First, iterative execution does not by itself establish agentic control: when program code determines whether to retry, re-prompt, or stop, the workflow remains agentless. Second, a system's self-description does not determine its classification. We examine the actual allocation of control rather than whether a study calls its components agents; where the cited account does not report the control flow in enough detail, we follow the authors' description and mark the entry as self-described. These distinctions allow us to compare systems with different implementations using a common architectural criterion.

\subsection{Agentless Workflows}
\label{sec:orch-agentless}

In agentless workflows, program code fixes what happens next and the LLM fills designated steps. This class covers a wide range of sophistication, from single-pass prompting to analysis-guided pipelines and iterative repair. The defining characteristic is not the simplicity of the workflow, but the retention of control by deterministic program logic.

\subsubsection{Single-Pass Prompting}

The simplest workflows issue one LLM call for each migration unit and evaluate the generated result externally. The model receives a predefined input and produces a target artifact, without selecting subsequent tools or deciding whether another operation is required. Although such workflows provide limited interaction with the development environment, they offer a straightforward baseline for evaluating the model's translation capability.

More elaborate translation sequences can also remain agentless when their execution is determined by an external algorithm. INTERTRANS, for example, leverages intermediate programming languages to improve code translation, but the translation paths are searched by the program rather than autonomously selected by the LLM~\cite{DBLP:conf/icse/Macedo00CA25}. Thus, multiple model calls and intermediate transformations do not by themselves establish agentic control. LAC2R uses Monte Carlo tree search to select model-proposed refinement actions and determine termination, so we code it as an algorithm-driven agentless workflow~\cite{DBLP:journals/corr/abs-2505-15858}.

\subsubsection{Analysis-Guided Pipelines}

A second group uses static analysis, dependency information, or predefined planning procedures to determine migration units, their processing order, and the context supplied to each LLM call. RustMap decomposes C projects according to usage dependencies and subsequently composes translated units with compiler and test feedback~\cite{DBLP:conf/iceccs/CaiLHYWLWYJ25}. AlphaTrans constructs a target skeleton and translates fragments in reverse call order before recomposing the repository~\cite{DBLP:journals/pacmse/IbrahimzadaKPAPSJ25}. C2SaferRust similarly combines program analysis with LLM-based rewriting of transpiler-generated code~\cite{DBLP:journals/tse/NitinKVR26}. In these systems, program analysis establishes the migration structure, while the LLM performs designated generation or refinement operations.

The same distinction applies to repository-level maintenance. Google's large-scale migration workflow combines change-location discovery with LLM-generated edits inside a scripted process~\cite{DBLP:conf/sigsoft/ZiftciNSKCK25}. CodePlan derives multi-file edit chains through incremental dependency and change-impact analysis~\cite{DBLP:journals/pacmse/BairiSKCIPRAS24}. Although these approaches incorporate planning and cross-file reasoning, the presence of a plan does not establish that the LLM controls its execution. When program code computes the plan and determines the subsequent operations, the workflow remains agentless.

These pipelines can support substantial repository-level migration without employing autonomous agents. Their main advantage is that program analysis provides explicit structural constraints and predictable execution. However, their ability to respond to unexpected migration conditions is bounded by the operations and decision rules encoded in advance.

\subsubsection{Scripted Repair Loops}

Many agentless workflows incorporate repeated generation and repair. Their distinguishing characteristic is that the triggers, retry conditions, and stopping rules are predefined by program code. C-to-Rust LLMigrate repeatedly prompts the model to repair translated functions until syntactic and compilation requirements are satisfied, while merging functions according to a predefined dependency order~\cite{DBLP:journals/corr/abs-2503-23791}. Byam combines compiler diagnostics and API differences to generate patches for breaking dependency updates and subsequently reruns the build or tests~\cite{DBLP:journals/ese/ReyesMBNBM26}.

These approaches demonstrate that feedback-driven refinement does not necessarily require an agent. The model may receive compiler or test diagnostics and generate corrective patches, while the surrounding program determines when to invoke the model again and whether the migration process should continue.

The distinction is particularly important when interpreting claims of autonomous repair. A workflow that repeatedly generates, compiles, and repairs code remains agentless if the retry policy and stopping condition are fixed in advance. Consequently, the presence of an iterative loop should not be used as sufficient evidence of agentic behavior.

\subsection{Single-Agent Migration}
\label{sec:orch-single}

In single-agent migration, one LLM instance selects which action to perform, what information to inspect, and when to stop. The surrounding harness defines the available tools and execution environment, but the agent determines how to use them. Consequently, reported results describe an agent--harness combination rather than the model alone.

\subsubsection{General-Purpose Coding Agents}

One approach applies existing coding agents to migration tasks without constructing a migration-specific agent architecture. Such agents operate on an existing repository, inspect relevant files, modify code, and use available development tools to obtain execution feedback.

Almeida et al. evaluate GitHub Copilot Agent Mode on automated library migration, examining its ability to perform changes across client applications~\cite{DBLP:conf/agent/AlmeidaVX26}. Unlike a single-pass translation workflow, the agent can interact with the repository and use the available development environment throughout the migration task.

Repository-level migration benchmarks provide additional settings for evaluating coding agents. FreshBrew evaluates AI agents on Java code migration~\cite{DBLP:journals/corr/abs-2510-04852}, TimeMachine-bench assesses model capabilities on repository-level migration tasks~\cite{DBLP:conf/eacl/FujiiMYS26}, and ScarfBench examines cross-framework application migration in enterprise Java~\cite{DBLP:journals/corr/abs-2605-06754}. MigrationBench also evaluates repository-level migration from Java 8~\cite{DBLP:conf/kdd/LiuLZCLNOSHTD26}. These benchmarks provide evidence about agent performance on migration tasks, although the benchmark results should not be interpreted as evidence that all evaluated agents employ the same internal architecture.

An important distinction is that the benchmark, the agent, and the execution harness represent different parts of the evaluation. The benchmark defines the migration tasks and acceptance conditions, while the agent determines how to perform the available operations. Therefore, results from different benchmarks cannot be used directly to establish the superiority of one agent architecture over another.

\subsubsection{Migration-Specific Agents}

Another approach constructs an action space around the requirements of a particular migration task. Rather than relying entirely on a predefined sequence, the agent selects operations according to the current migration state and feedback obtained from previous actions.

His2Trans describes an agentic C-to-Rust migration framework using historical migration knowledge~\cite{wang2026his2transknowledgeguidedagenticframework}, and BigBag describes its generation of transformation rules as agentic~\cite{DBLP:journals/corr/abs-2606-24446}. We code these two systems from their authors' description.

Migration-specific planning expands the available decision space. However, the autonomy of such systems remains constrained by the actions exposed through their harnesses. Their performance therefore depends on both the model's decision-making capability and the environment in which those decisions are executed.

Unlike agentless pipelines, single-agent systems can adapt their next action to intermediate observations. Nevertheless, this flexibility introduces additional dependencies on tool interfaces, context availability, and execution budgets. More autonomous decision-making does not necessarily guarantee more reliable migration results.

\subsection{Multi-Agent Migration}
\label{sec:orch-multi}

Multi-agent migration assigns distinct responsibilities to separate LLM instances, which exchange intermediate artifacts through messages or shared state. Unlike a single-agent system, the workflow distributes model-level responsibilities across multiple agents. We distinguish the resulting structures according to how responsibilities are separated and how intermediate results are passed between agents.

\subsubsection{Stage-Specialized Agent Chains}

The most common structure assigns one agent to each migration stage and passes intermediate artifacts along predefined hand-offs. RustPrint assigns planning, translation, synthesis, requirement refinement, test translation, and execution repair to six specialized agents. Its RequirementRefiner uses documentation differences to recover missing functionality~\cite{DBLP:journals/corr/abs-2605-14634}. RepoTransAgent separates repository retrieval, context collection, and refinement into specialized RAG, Context, and Refine agents~\cite{DBLP:journals/corr/abs-2508-17720}. ReCodeAgent distributes repository translation across analysis, planning, translation, and validation agents~\cite{DBLP:journals/corr/abs-2604-07341}. The Android-to-iOS pilot study similarly organizes specification extraction, code translation, and validation into separate agent responsibilities~\cite{DBLP:journals/corr/abs-2507-16037}. Mono2Sls assigns its pipeline to four role-specific agents with their own tools and a validation step~\cite{DBLP:journals/corr/abs-2604-24550}, and UITrans describes its UI translation as a collaboration of agents~\cite{DBLP:conf/internetware/GongWCHW25}.

These systems use role decomposition to separate different types of reasoning and intermediate context. For example, an analysis agent may construct project-level information, while a translation agent uses that information to produce target code and a validation agent examines the resulting artifacts.

However, the presence of specialized agents does not necessarily imply dynamic negotiation. When each agent receives a predefined task and passes its output to the next stage, the overall workflow remains a chain of fixed hand-offs. In such cases, multi-agent organization primarily separates responsibilities and contexts rather than distributing autonomous control over the overall migration process.

\subsubsection{Generator--Critic Structures}

A second topology separates artifact generation from the assessment of generated changes. ECAT employs a generator--discriminator workflow in which the generator proposes repository modifications and a discriminator evaluates them using a separate context~\cite{DBLP:journals/corr/abs-2608-09273}. The discriminator estimates code entropy and returns structured repair guidance, which the generator uses to revise the repository.

Unlike a simple sequential chain, this structure establishes an explicit feedback relationship between the component producing changes and the component assessing them. The separation of contexts is particularly relevant because a critic that shares the generator's assumptions may reproduce the same errors.

Nevertheless, an independent model-based critic is not equivalent to an independent correctness oracle. Its judgments remain subject to the limitations of the evaluation mechanism. The distinction between internal model feedback and external acceptance evidence is examined further in Section~\ref{sec:eval}.

\subsubsection{Deterministic Orchestrators over Specialized Agents}

Multi-agent systems may also place specialized agents under a program-controlled scheduler. In this arrangement, individual agents perform their assigned tasks, while deterministic code determines their execution order and coordinates the exchange of intermediate artifacts.

ORBIT illustrates this hybrid organization by constructing a dependency graph and target scaffold before coordinating specialized translation agents and deterministically verifying their outputs~\cite{DBLP:journals/corr/abs-2604-12048}. Although individual agents perform their assigned operations, the overall sequence remains controlled by program logic.

This structure remains multi-agent because multiple LLM instances assume distinct responsibilities and exchange intermediate results. However, its scheduling mechanism is deterministic rather than autonomously negotiated. Recording this distinction is important because the existence of multiple agents does not necessarily mean that control over their collaboration is distributed.

General multi-agent frameworks such as AutoGen, MetaGPT, and ChatDev provide interaction patterns for coordinating specialized LLM roles~\cite{wu2023autogenenablingnextgenllm,DBLP:conf/iclr/HongZCZCWZWYLZR24,DBLP:conf/acl/QianLLCDL0CSCXL24}. Their demonstrations offer useful architectural background, but they do not constitute direct evidence of migration effectiveness.

\subsection{Comparing Agent Architectures}
\label{subsec:agent-comparison}

Agent architectures are independent of the type of artifact produced by the LLM. A target implementation may be generated by a fixed pipeline, a single autonomous agent, or multiple cooperating agents. Similarly, refinement and transformation-rule synthesis can occur under different control structures. Table~\ref{tab:agent-architectures} compares the agent architectures of the 20 systems classified in this section in terms of control locus, the model roles examined in Section~\ref{sec:role}, and the evidence available to establish their structural contributions. Roles and architectures are shown in one table rather than cross-tabulated because they are linked by definition: every agentless system is Editor-only. Sixteen entries rest on a reported control flow and four on the authors' description. The table is not a census of Class~A: six of the remaining studies are benchmarks, and for 20 we did not code the structure.

\begin{table*}[t]
\caption{Orchestration structure and model role of the systems classified in
Sections~\ref{sec:role} and~\ref{sec:agent}. E: Editor; O: Orchestrator.
$\dagger$: coded from the authors' own description because the cited account
does not report the control flow in enough detail. NR: not reported in the
cited account.}
\label{tab:agent-architectures}
\centering\scriptsize
\setlength{\tabcolsep}{3pt}
\begin{tabular}{p{0.15\textwidth}p{0.19\textwidth}p{0.09\textwidth}p{0.25\textwidth}p{0.25\textwidth}}
\toprule
Structure and subtype & Who chooses the next step & Model role & Systems & Evidence about the structure \\
\midrule
\multicolumn{5}{l}{\emph{Agentless workflows (9 systems)}} \\
Algorithm-driven call sequence & Program (path search) & E &
  INTERTRANS~\cite{DBLP:conf/icse/Macedo00CA25} & NR \\
Algorithm-driven refinement search & Program (MCTS selection and stopping) & E &
  LAC2R~\cite{DBLP:journals/corr/abs-2505-15858} & NR \\
Analysis-guided pipeline & Program (static analysis or computed plan) & E &
  RustMap~\cite{DBLP:conf/iceccs/CaiLHYWLWYJ25},
  AlphaTrans~\cite{DBLP:journals/pacmse/IbrahimzadaKPAPSJ25},
  C2SaferRust~\cite{DBLP:journals/tse/NitinKVR26},
  Google~\cite{DBLP:conf/sigsoft/ZiftciNSKCK25},
  CodePlan~\cite{DBLP:journals/pacmse/BairiSKCIPRAS24} &
  CodePlan compares against baselines without planning; the others NR \\
Scripted repair loop & Program (fixed retry and stopping rule) & E &
  LLMigrate~\cite{DBLP:journals/corr/abs-2503-23791},
  Byam~\cite{DBLP:journals/ese/ReyesMBNBM26} & NR \\
\midrule
\multicolumn{5}{l}{\emph{Single-agent systems (3 systems, plus coding agents evaluated on benchmarks)}} \\
General-purpose coding agent & The agent, within the tools of its harness & E + O &
  Copilot Agent Mode~\cite{DBLP:conf/agent/AlmeidaVX26}; agents evaluated on
  FreshBrew, TimeMachine-bench, MigrationBench, and ScarfBench &
  Benchmark results vary agent, model, and harness together \\
Migration-specific agent & The agent & E + O &
  His2Trans$^\dagger$~\cite{wang2026his2transknowledgeguidedagenticframework},
  BigBag$^\dagger$~\cite{DBLP:journals/corr/abs-2606-24446} & NR \\
\midrule
\multicolumn{5}{l}{\emph{Multi-agent systems (8 systems)}} \\
Stage-specialized chain & Each agent within its stage; program code between stages &
  E + O, split across agents &
  RustPrint~\cite{DBLP:journals/corr/abs-2605-14634}, RepoTransAgent~\cite{DBLP:journals/corr/abs-2508-17720},
  ReCodeAgent~\cite{DBLP:journals/corr/abs-2604-07341},
  Mono2Sls~\cite{DBLP:journals/corr/abs-2604-24550},
  pilot study$^\dagger$~\cite{DBLP:journals/corr/abs-2507-16037},
  UITrans$^\dagger$~\cite{DBLP:conf/internetware/GongWCHW25} &
  RepoTransAgent's ``No Agent'' baseline removes role decomposition and
  refinement together; ReCodeAgent ablates individual agents and compares
  two single-agent variants; the others NR \\
Generator--critic & The critic sets repair targets; a programmed rule accepts or rejects changes &
  E (generator), O (critic) &
  ECAT~\cite{DBLP:journals/corr/abs-2608-09273} &
  Ablation without the discriminator (one agent generates and judges in a
  shared context); memory ablation \\
Program-scheduled agents & Program code schedules; agents act within assigned tasks &
  E + O & ORBIT~\cite{DBLP:journals/corr/abs-2604-12048} & NR \\
\bottomrule
\end{tabular}
\end{table*}

Three observations emerge from the comparison.

First, multi-agent migration systems predominantly employ stage-specialized agents with fixed hand-offs. RepoTransAgent and ReCodeAgent separate responsibilities across agents, while the Android-to-iOS pilot study organizes specification extraction, translation, and validation into successive stages~\cite{DBLP:journals/corr/abs-2508-17720,DBLP:journals/corr/abs-2604-07341,DBLP:journals/corr/abs-2507-16037}. ECAT instead establishes a generator--discriminator feedback relationship~\cite{DBLP:journals/corr/abs-2608-09273}. Although these structures differ in how they exchange intermediate results, the reviewed systems provide limited evidence of dynamic negotiation or autonomous reassignment of responsibilities among agents. Multi-agent organization therefore appears to emphasize the separation of responsibilities and contexts more than distributed decision-making.

Second, existing comparisons do not consistently isolate the contribution of agent architecture. A comparison between an agent-based system and a simpler baseline may simultaneously change the number of model calls, the availability of feedback, the division of responsibilities, and the execution budget. Consequently, an observed improvement cannot necessarily be attributed to the multi-agent structure itself. For example, RepoTransAgent's ``No Agent'' baseline removes both role decomposition and iterative refinement, making it difficult to isolate the contribution of agent collaboration~\cite{DBLP:journals/corr/abs-2508-17720}. ECAT does report a variant without the discriminator, in which a single agent generates and judges in a shared context: its model-judged score falls from 85.6 to 45.5 on Gallery, from 74.7 to 31.5 on AntennaPod, and from 71.2 to 12.0 on Meshtastic~\cite{DBLP:journals/corr/abs-2608-09273}. ReCodeAgent ablates individual agents and adds two single-agent variants built by condensing or concatenating the agents' prompts~\cite{DBLP:journals/corr/abs-2604-07341}. 
In these comparisons, however, roles, contexts, and prompts change together, equal total budget is not shown, and ECAT's outcome is a model's judgment rather than observed behavior. Structural evidence must therefore be distinguished from evidence that a particular system achieves a better migration result.

Third, greater autonomy does not necessarily imply better migration outcomes. Agentic systems may perform additional repository exploration, tool invocation, and iterative refinement, but these operations also consume time and computational resources. Multi-agent systems additionally incur communication and coordination overhead. Their benefits depend on whether the separation of responsibilities provides information or reasoning that a simpler workflow cannot obtain. Thus, comparisons should account for both acceptance outcomes and total execution cost.

Overall, the reviewed studies demonstrate several ways of organizing LLM-assisted migration, but provide limited controlled evidence about the independent contribution of agent architecture. A meaningful comparison would hold the model, migration inputs, acceptance oracle, and total execution budget constant while varying only the agent architecture. Such an evaluation would help distinguish improvements due to autonomous decision-making or agent collaboration from those arising from additional model calls, repair iterations, or computational resources.

%% file: sections/enabling-tech.tex
\section{Enabling Techniques}
\label{sec:technique}

While Sections~\ref{sec:se}--\ref{sec:agent} examine the activities supported by LLMs, the roles they assume, and how their work is organized, this section investigates the techniques that provide the knowledge, context, execution capabilities, and model capabilities required for migration (RQ4). We distinguish four complementary categories: Knowledge Engineering, Project Context, Environment \& Feedback, and Model Adaptation. They address different sources of migration difficulty: reusable knowledge about source--target transformations, information specific to the project being migrated, executable observations from the development environment, and adaptation of the model itself.

These techniques are not mutually exclusive and do not constitute a mandatory software stack. A migration system may combine a knowledge base with repository retrieval, compiler feedback, and a specialized training objective, while another may rely primarily on prompting and static analysis. We examine the artifacts and mechanisms used by each category, the migration problems they address, and the limitations of the available evidence.

\subsection{Knowledge Engineering}
\label{subsec:knowledge-engineering}

Knowledge engineering concerns the acquisition, representation, maintenance, and reuse of knowledge that supports migration across tasks or projects. Unlike project-specific context, which describes the particular source repository under migration, migration knowledge captures reusable information such as source--target API mappings, transformation rules, historical migration examples, known incompatibilities, and previously accepted repair decisions. Such knowledge can reduce the amount of migration reasoning that must be reconstructed independently for each task.

One form of migration knowledge is encoded in reusable transformation rules. Traditional approaches mine existing code changes to discover transformations that can be applied to other projects. MELT, for example, mines lightweight transformations from pull requests, using developer changes as evidence of recurring migration patterns~\cite{DBLP:conf/kbse/RamosMLMMG23}. More recent work investigates using LLMs to synthesize transformation rules directly from before-and-after examples. Allain et al. examine both the potential and limitations of LLM-generated code transformation rules, providing a mechanism for converting migration examples into reusable transformation knowledge~\cite{DBLP:journals/corr/abs-2609-03592}. These approaches differ in how the knowledge is obtained, but both aim to make migration decisions reusable beyond the original code change.

Another form preserves knowledge from previous migration experiences. His2Trans investigates a knowledge-guided agentic framework for project-level C-to-Rust migration, incorporating historical migration knowledge to support subsequent translation tasks~\cite{wang2026his2transknowledgeguidedagenticframework}. Rather than relying exclusively on the model's parametric knowledge, such approaches make migration-specific experience available as an explicit resource. This is particularly relevant when a project contains recurring patterns or when previous migrations have exposed incompatibilities that are not readily inferred from source code alone.

Migration knowledge can also be accumulated during execution. ECAT maintains a hierarchical memory tree containing accepted repair observations and retrieves relevant information in subsequent iterations~\cite{DBLP:journals/corr/abs-2608-09273}. This mechanism allows the system to reuse previously successful repair experience instead of repeatedly exploring the same problems. Its reported Gallery comparison provides evidence of memory use within that particular migration workflow. However, the admission criterion is based on code-entropy reduction rather than independent proof of behavioral correctness, and the result does not establish general memory effectiveness across different applications or ecosystems.

Memory and skills should also be distinguished. A memory item records an observation, decision, defect, or repair experience that may be useful later. A skill packages an operation with its inputs, preconditions, outputs, and checks, making a procedure reusable rather than merely preserving an individual experience. ECAT illustrates the use of repair and evaluation skills, but the available evidence does not isolate automatic skill discovery, cross-application reuse, or coverage maintenance across platform versions~\cite{DBLP:journals/corr/abs-2608-09273}.

The main challenge in knowledge engineering is therefore not simply accumulating more examples. Migration knowledge must retain its applicability conditions. An API mapping may depend on source and target SDK versions; a transformation rule may cover only particular usage patterns; and a previously accepted repair may be invalid under a different dependency configuration. Reusable knowledge should consequently preserve its source and target versions, preconditions, unsupported cases, and the checks that last validated it. The reviewed studies demonstrate several mechanisms for acquiring and reusing migration knowledge, but provide limited controlled evidence about its long-term accuracy and cross-project generalization.

\subsection{Project Context}
\label{subsec:project-context}

Project context engineering concerns the acquisition and representation of information specific to the source project being migrated. Migration decisions frequently depend on information beyond the immediately translated code unit, including project documentation, issue descriptions, architecture, dependencies, data and control flow, and relationships between files. This information differs from reusable migration knowledge: it describes the particular system and its implementation constraints rather than general transformation rules that can be transferred directly to other projects.

One approach retrieves existing project documentation and development records. The Android-to-iOS pilot study collects project documents and issue context to construct a knowledge base that supports subsequent translation~\cite{DBLP:journals/corr/abs-2507-16037}. Such information can help the model recover the purpose of source components and identify constraints that are not evident from individual methods. However, descriptive context does not necessarily specify which behaviors must be preserved or which platform-specific adaptations are acceptable. A project description may explain the intended functionality without providing testable acceptance conditions.

A second approach makes source architecture explicit through structured representations. RustPrint derives architecture documentation describing modules, data flows, APIs, and design rationales before C-to-Rust translation~\cite{DBLP:journals/corr/abs-2605-14634}. Such documentation provides a project-level blueprint that can guide the organization of migration work. Compared with general project descriptions, architectural documentation exposes relationships between components and makes structural dependencies easier to inspect. Nevertheless, a description of the source architecture does not automatically determine the appropriate target architecture, particularly when the target platform provides different abstractions or services.

Program structure and dependency representations provide another form of project context. C2RustXW encodes multi-level program structure and dependencies to guide project-level C-to-Rust translation~\cite{DBLP:journals/corr/abs-2603-28686}. PtrTrans preserves pointer relationships through a project knowledge graph, making relationships between program elements explicit during translation~\cite{DBLP:journals/pacmse/YuanMCSWLP26}. RustMap similarly decomposes projects according to usage dependencies and uses the resulting structure to organize translation and composition~\cite{DBLP:conf/iceccs/CaiLHYWLWYJ25}. These representations expose relationships that may be difficult to capture within a file-local prompt, including cross-file references, dependency ordering, and pointer usage.

Retrieval provides a complementary mechanism for supplying project context within the model's limited context window. Rather than placing an entire repository into a prompt, retrieval-based approaches select relevant files, code fragments, or documentation for the current task. Repository-level retrieval research, including RepoBench, evaluates retrieval and code completion in the presence of cross-file context~\cite{DBLP:conf/iclr/0003XM24}. Such resources provide useful enabling evidence, although repository retrieval performance alone does not establish that the retrieved information is sufficient for a migration task.

These approaches illustrate different representations of project context, from natural-language documentation to architecture descriptions and program graphs. They address different information requirements and should not be treated as interchangeable. Retrieved documentation may be irrelevant or contradictory; architecture descriptions may omit implementation details; and dependency graphs may preserve syntactic relationships without capturing platform-specific policies. Most importantly, none of these representations necessarily establishes the complete behavioral contract of the source application. Project context can reduce uncertainty during translation, but its completeness and consistency with the target requirements remain separate concerns.

\subsection{Environment \& Feedback}
\label{subsec:environment-feedback}

Environment and feedback engineering concerns the execution infrastructure and observations that allow a migration system to inspect, build, test, and refine its output. An execution environment provides more than a compiler interface: depending on the migration task, it may include a repository workspace, dependency installation, build configuration, deployment tools, device interaction, state initialization, and runtime observations. Feedback signals are the observations returned from these operations and used to determine whether a candidate should be modified or further examined.

Compiler and build feedback are among the most widely used signals in the reviewed migration studies. Byam collects compilation errors and API differences following dependency updates, uses them as context for LLM-generated patches, and subsequently reruns the build or tests~\cite{DBLP:journals/ese/ReyesMBNBM26}. C-to-Rust LLMigrate similarly uses compiler feedback to repair translated functions while assembling selected modules~\cite{DBLP:journals/corr/abs-2503-23791}. RustMap incorporates compiler and test feedback during project composition~\cite{DBLP:conf/iceccs/CaiLHYWLWYJ25}. These examples demonstrate how execution observations can support refinement beyond one-pass generation. However, a successful build establishes compilation and dependency consistency, not necessarily preservation of the application's behavior.

Test and differential-execution feedback provide additional evidence. C2RustXW combines compilation with differential execution to examine the behavior of translated programs~\cite{DBLP:journals/corr/abs-2603-28686}. Other migration workflows use supplied or generated tests to identify problems that are not exposed by compilation alone. GUPPY, for example, investigates deprecated Android API updates with generated-test feedback~\cite{DBLP:journals/tse/MahmudDCYNY26}. These signals can reveal behavioral differences within the exercised cases, although their usefulness depends on test coverage and the correctness of expected results.

Runtime interaction extends feedback beyond source-level checks. ECAT incorporates cold-start and smoke-click probes, as well as screenshot comparison, into its iterative repository migration process~\cite{DBLP:journals/corr/abs-2608-09273}. AndroidWorld provides an adjacent example of an execution environment with state initialization, emulator interaction, and state-based success checks for Android agents~\cite{DBLP:conf/iclr/RawlesCCWLFLBLC25}. AndroidWorld is not itself a cross-platform migration environment and does not provide a HarmonyOS or iOS acceptance oracle, but it illustrates the infrastructure needed to make application interactions observable and repeatable.

Repository migration benchmarks also make the relationship between environment and feedback explicit. TimeMachine-bench evaluates Python dependency and environment updates using executable tests, while FreshBrew evaluates Java and JDK migration with build and test-based acceptance conditions~\cite{DBLP:conf/eacl/FujiiMYS26,DBLP:journals/corr/abs-2510-04852}. These resources demonstrate the importance of reproducible environments and explicit acceptance checks. Nevertheless, their execution environments and acceptance criteria remain specific to their respective migration tasks and should not be interpreted as equivalent to complete mobile application validation.

Feedback must also be distinguished according to how it is used. When compiler diagnostics or test failures are fed back to the model to revise a candidate, they support the repair activity described in Section~\ref{sec:se-repair}. When tests are executed to determine whether the final result satisfies the migration contract, they serve as acceptance evidence. The same signal can participate in both processes, but its role depends on when and how it is used. In particular, an internal repair signal should not automatically be treated as an independent final evaluation.

For migration that must deliver a running application, the execution environment must expose operations and observations that reflect the target platform. A suitable harness should provide application installation and launch, interaction with relevant screens, state reset, logs, timeouts, and information about target services and device versions. Screenshots or UI hierarchy observations may further support visual and structural checks. These requirements are derived from the demands of application-level acceptance; they should not be interpreted as evidence that the reviewed systems already implement a complete cross-platform harness.

Environment and feedback engineering therefore determines which migration failures can be observed and acted upon. Compiler diagnostics are effective for syntactic and type errors, while tests, differential execution, and runtime probes expose different aspects of behavior. Their costs also differ, including environment setup, repeated execution, incomplete or flaky tests, and device-state management. The reviewed studies provide examples of these mechanisms at different migration granularities, but do not establish a common trade-off between their cost and behavioral coverage under a matched application-level workload.

\subsection{Model Adaptation}
\label{subsec:model-adaptation}

Model adaptation concerns changes to the model's parameters or training objectives to improve its performance on migration tasks. Unlike the preceding techniques, which primarily provide additional knowledge, project information, or execution feedback at inference time, model adaptation changes the model itself through task-specific training or optimization. Relevant dimensions include the training data, adaptation method, optimization objective, and the migration capability being targeted.

The general distinction is between inference-time support and parameter-level adaptation. Prompting, retrieval, and execution feedback can guide a pretrained model without modifying its parameters. Fine-tuning and reinforcement learning, in contrast, use task-specific examples or feedback objectives to update model behavior. This distinction is important because improvements obtained from additional context or repeated prompting should not automatically be attributed to model adaptation.

The reviewed migration literature contains comparatively limited evidence of explicit task-specific model adaptation. TRAVEL provides an example of reinforcement learning for function-level C-to-Rust translation, using compilation and test execution as training rewards to guide model optimization~\cite{DBLP:journals/corr/abs-2607-19966}. Unlike approaches that use compiler feedback only to repair an individual translation at inference time, this approach incorporates execution-based signals into the training objective. It therefore illustrates how migration-specific correctness signals can influence the model's learned behavior.

However, the scope of such evidence must be interpreted carefully. A training reward derived from compilation and tests reflects the properties measured by those signals. It does not automatically establish repository-level integration or application-level behavioral preservation. TRAVEL does compare adaptation with inference-time alternatives on one function-level task: with the same base model it reports computational accuracy of 38.83 for instruction prompting, 34.95 for in-context learning, 34.56 for retrieval augmentation, 68.35 for supervised fine-tuning, and 71.07 for its own method, and it ablates its components~\cite{DBLP:journals/corr/abs-2607-19966}. What the available evidence does not provide is such a comparison at the repository level, or one that matches the total cost of training and inference.

For future studies, reporting model adaptation requires more than identifying the base model and stating that it was fine-tuned. The training data should be characterized by its source, migration type, granularity, and target ecosystem. The optimization objective should specify whether it rewards syntactic validity, compilation, test execution, semantic preservation, or other migration properties. Evaluation should also separate training and held-out projects to establish whether adaptation generalizes beyond the examples used during training.

Overall, model adaptation represents a different route to improving migration capability from the inference-time techniques discussed above. Knowledge engineering supplies reusable migration experience, project context provides information about the current system, and environment feedback exposes execution outcomes. Model adaptation attempts to incorporate some of these capabilities into the model's learned behavior. However, explicit adaptation remains less extensively documented in the reviewed migration corpus than retrieval, structural guidance, and execution-based refinement. Its incremental value across migration types and granularities therefore remains an open empirical question.

The four categories address complementary sources of migration difficulty. Knowledge engineering makes migration experience reusable; project context makes repository-specific information accessible; environment and feedback engineering makes execution outcomes observable and actionable; and model adaptation changes the model's learned capabilities. They should not be treated as interchangeable mechanisms or counted as independent contributions without considering their interactions. A dependency graph may improve project understanding, but cannot establish behavioral correctness; execution feedback may repair a candidate, but cannot compensate for missing migration requirements; and a model trained on local translation rewards may still fail to integrate a complete application. The relevant question is therefore not how many techniques a system employs, but which migration constraints they address and what additional evidence they provide.

%% file: sections/evaluation.tex
\section{Evaluation Practices and Benchmark Resources}
\label{sec:eval}

Beyond the technical mechanisms that drive LLM-assisted migration, it is critical to understand how its success is rigorously measured. Accordingly, this section addresses RQ5 by investigating how these migration systems are evaluated and what their reported results empirically establish. We first examine the metrics used to accept a migrated result and the specific evidence each metric provides (Section~\ref{sec:eval-metrics}), followed by a review of the benchmarks available at each granularity level (Section~\ref{sec:eval-resources}). Finally, we discuss the protocol choices that determine whether an evaluation genuinely supports its underlying claims (Section~\ref{sec:eval-protocol}). Notably, unlike the testing activities described in Section~\ref{sec:se}---which a migration system performs internally to verify its own output---the evaluations discussed here refer to the external methodologies researchers use to assess the system as a whole.

\subsection{Metrics and What They Check}
\label{sec:eval-metrics}
\label{sec:eval-framework}

Migration studies report compilation rates, test pass rates, similarity
scores such as CodeBLEU, visual similarity, model-judged scores, and task
success rates. To compare them, we describe each check along four
dimensions and allow more than one label per dimension. The \emph{property}
is what the check is about: similarity to a reference or to the source,
build validity, idiomaticity or safety, visual correspondence, or functional
behavior (Table~\ref{tab:evidence-levels}). The \emph{scope} is how much of
the result the check covers: a unit, a repository, or a deployed application.
The \emph{means} is static when the migrated code is not run and dynamic when
it is. The \emph{judge} is what decides whether the check passed: a rule,
such as a compiler, a metric threshold, or a test assertion; a human; or a
model. The dimensions are separate. Executing tests is a means, not a
property, and it serves unit tests and end-to-end checks alike; the
difference between those two is one of scope. A model can judge any
property, by static or dynamic means. The properties provide complementary, scope-bounded evidence. Successful execution of a compiled application presupposes a valid build in that environment. Similarity alone leaves build validity and behavior unresolved, and successful compilation leaves behavior unresolved.

\begin{table*}[t]
\caption{Properties checked in migration evaluation. The properties are not ordered;
each establishes something different.}
\label{tab:evidence-levels}
\centering\small
\begin{tabular}{p{0.15\textwidth}p{0.22\textwidth}p{0.25\textwidth}p{0.09\textwidth}p{0.17\textwidth}}
\toprule
Property & Establishes & Does not establish & Usual means & Example use \\
\midrule
Similarity to a reference or the source & The output resembles a reference migration or the
  source in tokens or structure & That the code compiles, or behaves correctly & Static &
  CodeBLEU; structural alignment \\
Build validity & The code compiles, types check, and dependencies resolve &
  That the program launches or behaves correctly & Static & File or project compilation \\
Idiomaticity or safety & The output follows target-language practice on a proxy measure &
  That the code is correct, or maintainable in general & Static & Share of unsafe Rust code \\
Visual correspondence & The target UI resembles the source in appearance &
  That interactions, navigation, or state behave correctly & Static or dynamic &
  Page-level visual similarity \\
Functional behavior & The exercised cases produce the expected results, within the scope checked &
  Behavior outside the exercised cases; the adequacy of the tests & Dynamic &
  Unit and repository test suites; behavioral tests on a deployed application \\
\bottomrule
\end{tabular}
\end{table*}

Table~\ref{tab:metrics} lists the metrics used in the reviewed studies and
gives the property, means, and judge of each.

\begin{table*}[t]
\caption{Metrics used in migration evaluation, with the property they check, the means, and the
judge. The last column lists representative studies, not all users of a
metric.}
\label{tab:metrics}
\centering\footnotesize
\setlength{\tabcolsep}{1.7pt}
\begin{tabular}{p{0.16\textwidth}p{0.35\textwidth}p{0.18\textwidth}lll}
\toprule
Metric & Definition & Property & Means & Judge & Representative studies \\
\midrule
Exact match, BLEU, CodeBLEU & Lexical or structural similarity to a
  reference migration & Similarity & Static & Rule &
  \cite{DBLP:journals/corr/abs-2602-09930, DBLP:conf/nips/RoziereLCL20} \\
Structural alignment & Correspondence between source and target program
  elements & Similarity & Static & Rule &
  \cite{DBLP:journals/corr/abs-2608-09273} \\
Compilation success & Fraction of files or projects that compile on the
  target & Build validity & Static & Rule &
  \cite{zheng2026portingdeclarativeuiharmonyos} \\
Unsafe-code share & Share of unsafe code in translated Rust, as a proxy for
  idiomatic use of the target language & Idiomaticity or safety & Static & Rule &
  \cite{DBLP:journals/tse/NitinKVR26, DBLP:journals/pacmpl/EmreSDH21} \\
Visual similarity & Similarity between source and target UI renderings &
  Visual correspondence & Dynamic & Rule &
  \cite{zheng2026portingdeclarativeuiharmonyos} \\
Test pass rate, computational accuracy & Fraction of tasks or units whose
  tests pass on the target & Functional behavior & Dynamic & Rule &
  \cite{DBLP:conf/nips/RoziereLCL20, DBLP:journals/tse/WangWWGCGLMMZZ26, DBLP:journals/pacmse/IbrahimzadaKPAPSJ25, DBLP:journals/corr/abs-2604-07341, DBLP:conf/eacl/FujiiMYS26} \\
FreshBrew gate & The build succeeds, the original tests still pass, and line
  coverage drops by at most five points & Build validity; functional behavior & Dynamic & Rule &
  \cite{DBLP:journals/corr/abs-2510-04852} \\
MigrationBench gate & Build and tests succeed under the target JDK, compiled bytecode matches the target version, test-class and annotated-method name lists are preserved, and the executed test count does not decrease. Maximal migration additionally enforces the benchmark's dependency-version requirements. & Build validity; functional behavior & Dynamic & Rule &
  \cite{DBLP:conf/kdd/LiuLZCLNOSHTD26} \\
Model-judged score & A model rates whether requirements or semantics are
  preserved & Any & Either & Model &
  \cite{DBLP:journals/corr/abs-2608-09273} \\
Human inspection & Experts rate syntax, semantics, or completeness & Any & Either & Human &
  \cite{DBLP:journals/corr/abs-2507-16037} \\
Deployment and behavioral checks & The application deploys and passes
  behavioral tests & Build validity; functional behavior (deployed application) & Dynamic & Rule &
  \cite{DBLP:journals/corr/abs-2605-06754} \\
\bottomrule
\end{tabular}
\end{table*}

Metrics also differ in their unit and denominator. A fraction of compiled files, a proportion of
passed tests, a continuous score from a model judge, and a binary
application-level outcome measure different things, and the same word
``success'' can refer to any of them. Results from different benchmarks
therefore cannot be combined into a single migration success rate. RepoTransBench's
best success rate of
32.8\%~\cite{DBLP:journals/tse/WangWWGCGLMMZZ26}
and FreshBrew's best JDK~17 rate of
52.3\%~\cite{DBLP:journals/corr/abs-2510-04852},
for example, are not comparable, because they use different tasks and
acceptance criteria.

Similarity-based metrics, such as exact match, BLEU, and CodeBLEU, compare
the output with a reference migration and require no execution. They are
inexpensive and apply even to code that cannot yet be built, but a correct
migration can differ from the reference, and a similar one can still be
wrong. Computational accuracy was introduced for this reason, because BLEU
does not reflect whether a translation behaves
correctly~\cite{DBLP:conf/nips/RoziereLCL20}. Visual
similarity has an analogous limit: a page that resembles its source may still
differ in interaction, navigation, or state.

Execution-based metrics avoid this problem but introduce another: a test pass rate is only as informative as the tests behind it, which exposes it to two primary risks. 
First, weak test suites may fail to reject incorrect programs: augmenting benchmark tests frequently exposes flawed programs that the original tests had accepted~\cite{DBLP:conf/nips/LiuXW023}, and low coverage readily admits spurious solutions to migration tasks~\cite{DBLP:conf/eacl/FujiiMYS26}. Second, a system permitted to modify the tests can artificially achieve a passing result by deleting or weakening the assertions. Recent migration benchmarks therefore supplement test pass rates with safeguards on test identifiers, execution counts, or coverage. MigrationBench preserves test-class and annotated-method name lists and requires non-decreasing executed test counts~\cite{DBLP:conf/kdd/LiuLZCLNOSHTD26}, while FreshBrew limits line-coverage loss to five percentage points~\cite{DBLP:journals/corr/abs-2510-04852}. These safeguards strengthen evaluation, with behavioral assurance bounded by the checks performed and the cases exercised.
 
Beyond execution-based testing, model-based judgment introduces a different set of concerns. While an LLM acting as a judge can evaluate qualitative properties that are difficult to test automatically---such as whether a high-level requirement is met---its alignment with independent correctness checks must be rigorously measured rather than merely assumed. Self-review poses a particularly severe risk: a model tasked with reviewing its own migrated code may articulate its reasoning fluently while completely overlooking subtle semantic errors~\cite{DBLP:journals/corr/abs-2605-21537}.
\subsection{Benchmarks and Empirical Resources}
\label{sec:eval-resources}

Table~\ref{tab:eval-resources} groups the migration benchmarks by the
granularity levels of Section~\ref{sec:bg-types}. The figures follow the
cited versions.

\begin{table*}[t]
\caption{Migration benchmarks grouped by target granularity. The last column gives the property checked, the means, and the judge (Table~\ref{tab:evidence-levels}). $\ast$: the task requires a deployed application.}
\label{tab:eval-resources}
\centering\footnotesize
\setlength{\tabcolsep}{2.5pt}
\begin{tabular}{p{0.07\textwidth}p{0.14\textwidth}p{0.18\textwidth}p{0.17\textwidth}p{0.22\textwidth}p{0.14\textwidth}}
\toprule
Level & Resource & Migration task & Scale & Acceptance criteria & Property; means; judge \\
\midrule
Module & JMigBench~\cite{DBLP:journals/corr/abs-2602-09930} &
  Java~8 $\rightarrow$ Java~11 function updates & Function-level tasks across
  deprecated-API categories & CodeBLEU, keyword checks, exact matches &
  Similarity; static; rule \\
 & ArkTrans dataset~\cite{zheng2026portingdeclarativeuiharmonyos} &
  Compose or SwiftUI $\rightarrow$ ArkUI pages & 100 parallel pages; 75
  evaluation files per direction & File compilation; global and local visual
  measures; external dependencies excluded & Build validity, visual; static compilation and dynamic rendering; rule \\
 & Python~2 snippets~\cite{DBLP:journals/corr/abs-2605-21537} &
  Python~2 $\rightarrow$ Python~3 modernization & 60 snippets &
  Behavior checked by a type-strict oracle & Behavior; dynamic; rule \\
\midrule
Repository & RepoTransBench~\cite{DBLP:journals/tse/WangWWGCGLMMZZ26} &
  Cross-language repository translation & 1,897 repositories; 13 language
  pairs & Executable test suites & Behavior; dynamic; rule \\
 & MigrationBench~\cite{DBLP:conf/kdd/LiuLZCLNOSHTD26} &
  Java~8 $\rightarrow$ Java~17 in the reported experiments (21 also supported) & 5,102 repositories in the full set;
  300 in the selected subset & Build and tests, target bytecode version, preserved test-class/method name lists, and a non-decreasing executed test count. The maximal tier additionally enforces dependency-version requirements. & Build validity, behavior; dynamic; rule \\
 & FreshBrew~\cite{DBLP:journals/corr/abs-2510-04852} &
  Java $\rightarrow$ JDK~17 & 228 Maven repositories & Build, original tests
  still passing, coverage drop within five points & Build validity, behavior; dynamic; rule \\
 & TimeMachine-bench~\cite{DBLP:conf/eacl/FujiiMYS26} &
  Python dependency updates & 1,145 Full repositories; 100 Verified repositories & Previously failing tests pass & Behavior; dynamic; rule \\
 & A2H-RepoBench$^{\ast}$~\cite{DBLP:journals/corr/abs-2608-09273} &
  Android $\rightarrow$ HarmonyOS repositories & Three repositories with
  scaffold and specifications & Structural alignment and model judgment &
  Similarity; static; rule and model \\
 & ScarfBench$^{\ast}$~\cite{DBLP:journals/corr/abs-2605-06754} &
  Spring, Jakarta~EE, and Quarkus applications & 34 application families; 102 variants; 204 directed tasks & Compilation, container deployment, held-out behavioral
  tests & Build validity, behavior on the deployed application; dynamic; rule \\
\bottomrule
\end{tabular}
\end{table*}

Evaluation sets built for a single system, such as the applications on which Mono2Sls reports cloud deployment and behavioral tests~\cite{DBLP:journals/corr/abs-2604-24550}, are not listed.

Three patterns emerge from the table. First, what is checked differs by
granularity: module-level resources rely mainly on static and visual checks,
repository-level resources mostly on test suites, and ScarfBench combines deployment with execution-based behavioral acceptance. Second, the resources differ
in migration trigger, language ecosystem, task unit, and acceptance criteria,
so each defines its own notion of success. Third, mobile migration is covered at the module level by page-level UI
datasets and at the repository level by A2H-RepoBench, which evaluates
Android-to-HarmonyOS repositories through structural alignment and
model-based
judgment~\cite{DBLP:journals/corr/abs-2608-09273}.
A benchmark whose acceptance criterion is independent, execution-based behavior of the migrated mobile application on the target platform
has not yet appeared among the resources we reviewed and would be a natural
next step.

While comprehensive benchmarks for application-level migration do not yet exist, adjacent research areas offer valuable building blocks for constructing them. For instance, general code generation benchmarks have already established robust methods for evaluating function synthesis via execution~\cite{DBLP:journals/corr/abs-2107-03374}, scaling these evaluations across multiple programming languages~\cite{DBLP:journals/tse/CassanoGNNPPYZAFGGJ23}, and assessing repository-level edits that resolve real-world issues~\cite{DBLP:conf/iclr/JimenezYWYPPN24}. In the mobile domain, environments like AndroidWorld can initialize device states and verify task outcomes across numerous apps~\cite{DBLP:conf/iclr/RawlesCCWLFLBLC25}, while AppEval validates automated repairs by running tests directly on installed applications~\cite{DBLP:journals/corr/abs-2608-18588}. Although none of these frameworks were designed specifically for software migration, they collectively demonstrate that automated, execution-based evaluation on fully installed applications is technically feasible. These methodologies could readily be adapted to evaluate paired source and target applications in future migration benchmarks.

\subsection{Protocols and Credibility}
\label{sec:eval-protocol}

Beyond the choice of benchmark and metric, two aspects of the experimental
protocol most directly determine what a reported result can support.

\paragraph{Independence of feedback and acceptance}
Many migration systems run tests during repair (Section~\ref{sec:se}). If the
same tests are then used to report the final result, the reported pass rate
measures how well the system fits those tests rather than how well it
preserves behavior. Acceptance checks should therefore be held out from the
feedback loop, or the overlap should be reported. ScarfBench uses a hidden containerized harness to rebuild submissions and execute behavioral tests~\cite{DBLP:journals/corr/abs-2605-06754}.

\paragraph{Inputs and human participation}
Migration tasks vary significantly in the inputs provided to the system. Source-only conversion is inherently more challenging than tasks supplemented with formal specifications, target scaffolds, or few-shot examples. For instance, ECAT initiates the migration from a predefined target template~\cite{DBLP:journals/corr/abs-2608-09273}, whereas ArkTrans dynamically constructs a target skeleton and leverages explicit demonstrations~\cite{zheng2026portingdeclarativeuiharmonyos}. Human involvement also extends to task curation---such as verifying that benchmark tasks are solvable~\cite{DBLP:conf/eacl/FujiiMYS26}---as well as to manually completing migrations or selecting among generated outputs. Because this manual effort constitutes a substantial portion of the overall migration cost, evaluations must transparently report which artifacts were prepared by hand and exactly which inputs were accessible to the system.

Three additional methodological aspects warrant consideration. First, baselines only substantiate claims about a specific mechanism if they isolate that variable while holding all other factors constant
(Section~\ref{sec:challenge-attribution}). Second, the risk of data contamination---where benchmark repositories inadvertently appear in a model's training corpus---must be mitigated, for example, through time-indexed task
collection~\cite{DBLP:conf/iclr/JainHGLYZWSSS25}.
Finally, to ensure fair comparison and reproducibility, comprehensive cost reporting must encompass both failed computational attempts and the aforementioned manual overhead, while experimental logs should meticulously record the specific model, prompt, and SDK versions utilized
(Section~\ref{sec:challenge-cost}).

In summary, our investigation into migration evaluation (RQ5) reveals that current methodologies substantiate only bounded claims. Specifically, repository-level test suites can only verify that explicitly exercised behaviors are preserved; page-level checks merely confirm that UI code compiles and visually resembles the source; and structural or model-based scores serve as weak proxies for functional correctness. Consequently, these disparate metrics preclude the calculation of a unified, end-to-end migration success rate. Most notably, within our reviewed corpus, independent execution-based acceptance of a migrated mobile application against a specified behavior contract remains absent; ECAT's emulator probes feed repair rather than acceptance. Bridging this critical evaluation gap will require the development of benchmarks featuring paired source and target applications, the adoption of acceptance checks strictly independent of the system's internal feedback loops, and the transparent reporting of all manual overhead and total costs. Section~\ref{sec:challenges} synthesizes these pressing methodological needs into a broader discussion of open problems.

%% file: sections/challenges.tex
\section{Challenges and Open Problems}
\label{sec:challenges}

The preceding analyses demonstrate substantial progress in module-level migration and a
rapidly growing body of work at the repository level; however, the path toward full
application-level integration remains a largely unresolved challenge. This section discusses
six open problems that arise from these analyses
(Sections~\ref{sec:se}--\ref{sec:eval}). For each problem, we summarize the
current empirical evidence, explain why it falls short, and outline a strategic research
direction. Table~\ref{tab:challenge-agenda} summarizes the resulting research agenda.

\subsection{From Module Success to Application Integration}
\label{sec:challenge-integration}

Most contemporary methods transform and evaluate units in strict isolation. UI migration
studies assess independent pages~\cite{zheng2026portingdeclarativeuiharmonyos} or assemble a project's UI without showing that the behavior behind it is
preserved~\cite{DBLP:conf/internetware/GongWCHW25},
and even repository-level techniques succeed on only a minority of complex
tasks~\cite{DBLP:journals/tse/WangWWGCGLMMZZ26}.
Building on the reported mobile-development challenges~\cite{DBLP:conf/esem/JoorabchiMK13}, we identify cohesive navigation, event handling, permissions, persistence, and background tasks as application-level integration requirements. Pages that function correctly in isolation can still trigger data loss when the underlying platform destroys and
recreates them during lifecycle
transitions~\cite{DBLP:conf/issta/RiganelliMRMM20}.

Dependencies among functions exacerbate this gap. If a login module is migrated
incorrectly, users are prevented from accessing their orders or editing their profiles, even
if those downstream functions are migrated flawlessly. A function can therefore be technically
correct in isolation yet entirely unreachable in the migrated application; thus, localized
evaluations provide little assurance regarding whether users can actually complete their tasks. Treating
every unreachable function as defective blurs the origin of the fault, whereas
omitting unreachable functions from the evaluation artificially overstates usability.

A promising direction is to formalize explicit cross-module contracts encompassing
dependencies, events, shared state, and resources, utilizing them for both
planning and validation. Two questions naturally follow: Do such contracts improve
application-level outcomes when the generator, inputs, and computational budget remain fixed? And do module-level scores reliably predict the success of complete user journeys? Answering these requires evaluations that precisely map functional dependencies, assessing both individual functions under valid preconditions and complete user journeys originating from realistic initial states.

\subsection{Platform Differences Without Direct Counterparts}
\label{sec:challenge-platform}

Cross-platform migration frequently encounters capabilities lacking direct one-to-one counterparts on the target platform, such as distinct permission models, background
execution policies, and system services. This discrepancy leads to two types of errors. An API
replacement may compile successfully yet inadvertently alter observable timing, permissions, or
persistence behaviors. Conversely, the absence of an identical target API
does not inherently render the user's goal infeasible, as an alternative interaction paradigm or
architectural pattern might achieve the same outcome. An Android-to-iOS study reported translation failures at the
method, file, and package levels even when relevant project context was
retrieved~\cite{DBLP:journals/corr/abs-2507-16037}.

The platform migration systems we reviewed largely rely on the assumption of literal correspondence: they map each source
construct to a target construct and several of their checks compare the result with the
source. Apart from individual allowances (Section~\ref{sec:se-spec}), decisions regarding which deviations are acceptable are left implicit,
usually delegated to human developers inspecting the final output. While this assumption holds for
language translation within a single ecosystem, it fails entirely when the underlying platform
architecture changes.

Migration paradigms therefore require specifications that disentangle three distinct elements: the
outcomes that must be strictly preserved, the platform-specific variations that are deemed
acceptable, and the capabilities for which no viable target behavior currently exists. For example, a confirmed edit may need to survive a process restart,
without requiring the target platform to mimic the source application's screen transitions perfectly. LLMs
could assist in recovering such nuanced specifications from source code and documentation,
directly linking this challenge to the specification activities discussed in
Section~\ref{sec:se}. If a requirement is relaxed following a migration failure, this adjustment
should be transparently reported as a scope change, rather than being counted as a successful migration.

\subsection{Non-Functional Properties of the Migrated System}
\label{sec:challenge-nonfunctional}

The evaluations in Section~\ref{sec:eval} focus almost exclusively on functional behavior. 
In Table~\ref{tab:metrics}, only the share of \texttt{unsafe} Rust code addresses a non-functional property, and merely as a static proxy for memory safety~\cite{DBLP:journals/tse/NitinKVR26, DBLP:journals/pacmpl/EmreSDH21}; no benchmark in Table~\ref{tab:eval-resources} measures latency, memory, energy, binary size, security, or accessibility (maintainability is discussed in Section~\ref{sec:challenge-cost}). 
Yet non-functional goals (e.g., memory safety or operational elasticity) often motivate migration in the first place. 
Worse, literal translations and repair loops rewarded solely for passing tests can silently introduce regressions---such as redundant copies, blocking calls, or \texttt{unsafe} blocks added merely to appease compilers.

For mobile migration, we propose checking startup latency, rendering smoothness, energy use, permission scope, and lifecycle state preservation. These checks extend the reported mobile-development and lifecycle concerns~\cite{DBLP:conf/esem/JoorabchiMK13,DBLP:conf/issta/RiganelliMRMM20} across fragmented OS versions and devices~\cite{DBLP:conf/kbse/WeiLC16,DBLP:journals/ese/ScalabrinoBLPLO20}. 
Validating these properties is inherently difficult: they lack binary pass-or-fail oracles, suffer from hardware- and workload-dependent measurement noise, and trade off against one another and functional completeness.

We therefore advocate defining explicit non-functional budgets relative to the source system---profiled across repeated trials on identical devices and workloads---and reporting them alongside functional outcomes rather than within a single score. 
Three open questions follow: how often functionally accepted migrations meet these budgets, whether profiling-guided repair loops can mitigate regressions, and which properties can be estimated statically before the target builds.
\subsection{Generalization Across Directions, Ecosystems, and Versions}
\label{sec:challenge-generalization}

Current empirical evidence is disproportionately concentrated in a few specific migration directions---notably C to
Rust (14 of the 46 direct migration studies) and version and API upgrades (17), whereas platform migration is represented by only five studies---with evaluations typically relying
on small sample sizes. ECAT covers only a single direction across a limited set of
repositories~\cite{DBLP:journals/corr/abs-2608-09273},
and page-level UI migration restricts the dependencies of the evaluated
pages~\cite{zheng2026portingdeclarativeuiharmonyos}.
Applications heavily dependent on hardware, payment gateways, or third-party
services are rarely investigated. Earlier porting efforts address specific source--target platform pairs~\cite{DBLP:journals/mis/LamhaddabLE19,DBLP:journals/monet/PuderA13}. We therefore treat transfer across migration directions as an empirical question.

Platform evolution introduces a secondary source of distribution shift. Mobile platform APIs evolve
rapidly while client adoption inevitably lags
behind~\cite{DBLP:conf/icsm/McDonnellRK13},
and compatibility issues across different OS versions and devices are
notoriously common~\cite{DBLP:conf/kbse/WeiLC16, DBLP:journals/ese/ScalabrinoBLPLO20}.
Consequently, mapping knowledge collected for a specific SDK version can rapidly become obsolete,
and a model trained prior to a major release remains ignorant of newly introduced APIs. Furthermore, the inclusion of open-source benchmark applications in pretraining data
artificially inflates performance metrics on familiar codebases.

Future benchmarks must therefore strictly hold out entire application families, migration
directions, and SDK versions, reporting results separately for
applications with varying hardware and service dependencies. Versioned
knowledge bases present a concrete intervention to evaluate: comparing the performance of current versus
stale mapping snapshots on held-out platform versions would reveal the extent to which
migration quality depends on perfectly up-to-date domain knowledge.

\subsection{Attributing Gains to Roles, Structures, and Techniques}
\label{sec:challenge-attribution}

Migration systems typically fuse a model role, an orchestration structure, and
several enabling techniques; however, studies rarely isolate the individual contributions of these components. For instance, RepoTransAgent's baseline strips away both the role decomposition and
the iterative refinement loop, making it impossible to attribute the performance gains strictly to its multi-agent
architecture~\cite{DBLP:journals/corr/abs-2508-17720}.
ECAT and ReCodeAgent do report structural comparisons, namely a variant without the discriminator and two single-agent
variants, respectively~\cite{DBLP:journals/corr/abs-2608-09273, DBLP:journals/corr/abs-2604-07341}, but in both the roles, contexts, and prompts change together and matched total cost is not shown (Section~\ref{subsec:agent-comparison}).
Because models, tasks, acceptance criteria, and computational budgets vary wildly across studies,
direct cross-study comparisons are currently infeasible.

\begin{table*}[t]
\caption{Research agenda derived from the identified open problems. The suggested studies
represent proposals generated by this survey.}
\label{tab:challenge-agenda}
\centering\small
\begin{tabular}{p{0.17\textwidth}p{0.35\textwidth}p{0.40\textwidth}}
\toprule
Problem & Open question & Suggested study \\
\midrule
Integration (\S\ref{sec:challenge-integration}) & Do explicit cross-module
  contracts improve application-level outcomes, and do module-level scores
  predict them? & Same application pairs evaluated with and without contracts; fixed
  generator, inputs, and budget; outcomes measured via complete user
  journeys \\
Platform differences (\S\ref{sec:challenge-platform}) & Can intent-level
  specifications accept valid platform adaptations without inadvertently accepting lost
  functionality? & Acceptable variations strictly annotated in advance; comparison of intent-level
  versus literal correspondence; tracking rates of false acceptance and false
  rejection \\
Non-functional properties (\S\ref{sec:challenge-nonfunctional}) & How often do functionally accepted migrations meet source-relative budgets, and can profiling-guided repair reduce regressions? & Repeated profiling on matched devices and workloads; latency, memory, energy, and security budgets reported alongside functional outcomes \\
Generalization (\S\ref{sec:challenge-generalization}) & Do performance gains persist on
  unseen applications, migration directions, and SDK versions? & Strict holdouts split by
  application family and version; comparison of current versus stale mapping knowledge;
  results stratified by hardware and external service dependencies \\
Attribution (\S\ref{sec:challenge-attribution}) & Does the orchestration
  structure or the assigned model role independently improve results? & Controlled studies varying one factor at a
  time; fixed generator, inputs, acceptance criteria, and total budget, with token-matched and time-matched designs reported separately \\
Cost and maintenance (\S\ref{sec:challenge-cost}) & What is the true total cost of an accepted
  migration, and how expensive is the resulting code to maintain? &
  Budget sweeps accounting for failures and human labor; longitudinal comparisons
  against manual reimplementation \\
\bottomrule
\end{tabular}
\end{table*}

Consequently, it remains ambiguous whether reported performance gains stem from the orchestration
structure itself or merely from an increased number of model calls and repair iterations. The
same ambiguity applies to model roles: it remains unmeasured whether an LLM tasked with operating build and
analysis tools genuinely provides added value over a deterministic script invoking those exact same tools.

Addressing these ambiguities necessitates controlled ablation studies that isolate single variables,
such as the orchestration structure, the model role, or a specific prompting technique, while
strictly holding the generator, inputs, acceptance criteria, and total computational
budget constant; because equal tokens do not imply equal time or cost, token-matched and time-matched designs should be reported separately. The adoption of shared evaluation harnesses and the transparent reporting of negative results would significantly enhance the reproducibility of such comparisons.

\subsection{Cost, Maintenance, and Evolution}
\label{sec:challenge-cost}

Reporting of migration costs remains highly inconsistent. While some studies report token usage or iteration counts---and
search-based planning is widely acknowledged to inflate computational
costs~\cite{DBLP:journals/corr/abs-2505-15858, DBLP:journals/corr/abs-2608-09273}---metrics regarding
failed attempts, evaluation infrastructure overhead, and the manual human labor required to
finalize a migration are rarely quantified. At Google, developers estimated that
LLM assistance effectively halved migration
time~\cite{DBLP:conf/sigsoft/ZiftciNSKCK25}; conversely, a randomized
study of general software development, not of migration, revealed that experienced developers actually took longer when utilizing
AI assistance, despite subjectively anticipating a speed
increase~\cite{DBLP:journals/corr/abs-2507-09089}.
Consequently, perceived productivity and objectively measured efficiency can diverge significantly.

Post-migration maintenance remains critically underexplored. Developers working on
the target platform inherit the responsibility of maintaining the migrated code, yet its idiomaticity,
readability, and the long-term cost of future modifications are seldom evaluated. Furthermore, the migration
toolchain itself evolves: models are continuously retired and SDKs updated. Reusable
transformation rules present a similar vulnerability; a rule that syntactically applies to
a snippet of code is not guaranteed to be semantically correct for
it~\cite{DBLP:journals/corr/abs-2609-03592},
and rules synthesized for one SDK version may fail on subsequent codebase iterations.
Thus, a migration capability demonstrated at one point in time possesses a strictly limited shelf life.

Future studies must report the comprehensive cost per accepted migration across all attempts,
explicitly accounting for failures and manual completion, and plot quality-cost trade-offs
by systematically varying the token budget. Longitudinal studies should compare LLM-migrated code against
manually reimplemented code throughout subsequent requirement changes and platform updates.
Meticulously logging the exact models, prompts, tools, knowledge snapshots, and SDK versions utilized
in each experiment is essential to ensure results remain reproducible when the underlying ecosystem evolves.


%% file: sections/conclusion.tex
\section{Conclusion}
\label{sec:conclusion}

This survey has reviewed LLM-assisted software migration across languages, versions, APIs, frameworks, and platforms, tracing the progression from module-level transformation to repository-level integration and complete application delivery. 
Analyzing the literature across software engineering activities, model roles, orchestration structures, and enabling techniques alongside current benchmarks yields four principal conclusions. 
First, although 30 of the 46 direct migration studies target entire repositories---including four whose tasks require application deployment---ensuring that migrated components operate cohesively in the target environment remains challenging. 
Second, progress across the lifecycle is uneven: software construction is well supported, whereas requirements engineering remains underdeveloped, with most systems treating legacy code and existing tests as the sole specification and rarely testing deployed applications. 
Third, while complex systems increasingly rely on tool-using models within multi-agent workflows, few studies isolate the gains of these architectures under controlled conditions. 
Fourth, current evaluations check distinct properties---from compilation and visual similarity to executed tests and runtime behavior---that cannot be combined into a single success rate; notably, no benchmark in our corpus evaluates migrated mobile applications via independent, execution-based acceptance against a target-platform behavior contract.

Advancing toward true application-level migration therefore requires specifications that distinguish preserved behaviors from permissible platform adaptations, native target-platform acceptance checks kept strictly independent of internal repair loops, benchmarks featuring paired source and target applications, and controlled studies reporting total migration cost, including human intervention and long-term maintenance. 
Together with the research agenda in Section~\ref{sec:challenges}, this framework aims to shift the measure of progress toward whether migrated applications reliably serve their end users.